\documentclass[twocolumn,twocolappendix]{aastex701}
\shorttitle{Data-driven RMHD Simulations of AR 11158 and the X class flare}
\shortauthors{F. Chen}
\usepackage{amsmath}
\usepackage{color}
\newcommand{\sectref}[1]{Section\,\ref{#1}}

\newcommand{\figref}[1]{Figure\,\ref{#1}}
\newcommand{\figsref}[1]{Figures\,\ref{#1}}
\newcommand{\tabref}[1]{Table\,\ref{#1}}

\newcommand{\appdref}[1]{Appendix\,\ref{#1}}

\begin{document}

\title{Data-driven Radiative Magnetohydrodynamics Simulations with the MURaM Code: \\
the Emergence of Active Region 11158 and the X2.2 Flare}

\correspondingauthor{Feng Chen}
\email{chenfeng@nju.edu.cn}

\author[0000-0002-1963-5319]{Feng Chen}
\email{chenfeng@nju.edu.cn}
\affiliation{School of Astronomy and Space Science, Nanjing University, Nanjing 210023, China}
\affiliation{Key Laboratory of Modern Astronomy and Astrophysics (Nanjing University), Ministry of Education, Nanjing 210023, China}

\begin{abstract}
We present the application of the data-driven branch of the MURaM code to the extensively studied flare-productive active region 11158. We refine the hybrid model strategy, which was described in the earlier paper of this series, to model the emergence of the active region during 4 solar days starting shortly before 2011 February 11 and the eruption of an X2.2 flare on February 15. After 4 days of evolution, a major eruption of a magnetic flux rope occurs in the simulation at approximately 3 hours (3\% difference) before the real flare. The eruption leads to magnetic reconnection, which contributes to bulk heating in the chromosphere and corona. The deposition of flare energy in the chromosphere causes strong condensations and evaporations, which fill hot post-flare loops and bright flare ribbons that exhibit separation and extension similar to the observed ribbon evolution. The synthesized soft X-ray flux corresponds to X class, which is close to the real event. The upward eruption of the flux rope leads to a piston-driven shock and horizontal expansion that exerts a strong downward impact on the lower atmosphere and generate an apparently fast-propagating chromospheric Moreton wave. We conclude that the data-driven radiative simulation of this active region can reproduce the key observational results of the real flare and demonstrate the great potential of this method for studying solar eruptions in a realistic corona environment.
\end{abstract}


\section{Introduction} \label{sec:intro}
Solar eruptions represent perhaps the most complex and dynamic interaction between the solar magnetic field and plasma. The community has recognized the important application of numerical simulations that follow the observed evolution of photospheric magnetic fields and consider sophisticated physical processes for realistic plasma properties.\citep{Patsourakos+al:2020,JiangChaowei+al:2022,GuoYang+al:2024,JiangChaowei:2024,Schmieder+al:2024}

Following previous papers in this series that described the implementation of the data-driven boundary in the Radiative MHD code MURaM \citep{Chen+al:2023} and the application on non-eruptive emerging an active region \citep{Chen:2025}, we apply this method to a well-known and extensively studied active region 11158, which produced the first X class flare in solar cycle 24 \citep{Schrijver+al:2011} and has been thoroughly analyzed from its emission properties \citep[e.g., ][]{Young+al:2013,Milligan+al:2014} to magnetic structures \citep[e.g., ][]{Liu+Schuck:2012,SunXudong+al:2012,WangShuo+al:2012,Vemareddy+al:2012,Tziotziou+al:2013,Kazachenko+al:2015,SunXudong+al:2017}.

In particular, data-driven magnetohydrodynamic (MHD) simulations under different approximations have been employed to investigate the evolution of and eruptions in this active region. \citet{Cheung+DeRosa:2012} and \citet{Lumme+al:2017} conducted a magneto-frictional calculation to model the evolution of the coronal magnetic field over several days of emergence. These models omit the role of plasmas and represent the progress of methodology from a series of static magnetic field calculations \citep{Jarolim+al:2023} to a full MHD simulation that includes both magnetic fields and plasmas. 

\citet{Inoue+al:2014,Inoue+al:2015} used a nonlinear-forcefree-field as the initial condition for an MHD simulation and modeled the magnetic evolution during the X class flare, and \citet{Inoue+al:2018} investigated the M6.6 flare that occurred two days earlier. \citet{Hayashi+al:2018,Hayashi+al:2019} conducted MHD simulations for approximately 15 hours with a focus on testing different methods of driving the boundary and the accumulation of magnetic energy in the coronal volume.

Recently, MHD simulations with improved treatments for plasma thermodynamics have become possible. The hybrid model of \citet{Afanasyev+al:2023} covers the emergence of the magnetic field via a data-driven magneto-friction calculation and the eruption via a data-constrained MHD simulation with thermal conduction and an almost isothermal equation of state. \citet{Fan+al:2024} improved the energy equation by including the missing radiative loss and more self-consistent heating to maintain the corona temperature. Their model setup has a sufficient spatial extent, which allows investigations of the large-scale impact of the eruption, such as coronal extreme-ultraviolet (EUV) waves. A global scale model of the X class flare in AR 11158 was reported by \citet{Jin+al:2022}, who investigated the large-scale dimming associated with coronal mass ejection (CME). \citet{Hoeksema+al:2020} (and references therein) presented a comprehensive test of the capability of driving local/global scale magneto-frictional models and radiative MHD models that reproduced the pre-eruption state of AR 11158.

The well-studied AR 11158 serves as an ideal example of testing and validating the data-driven branch of the MURaM code, which may provide new insight complementing the detailed studies that have been performed. This work is a part of a comprehensive project of an international collaboration to model evolving and erupting active regions. As the first attempt, we prefer to use a simple setup following \citet{Cheung+DeRosa:2012} to drive the evolution, which is sufficient to capture the most important magnetic structure for the eruption. The comprehensive model describes, at the same time, the eruption at the large-scale corona, as well as the detailed plasma dynamics in different layers of the atmosphere, which, to the best of our knowledge, has not been done in previous works. The rest of the paper is organized as follows. We described in \sectref{sec:method} the multi-stage model strategy and setup of the numerical simulations. The results are presented in \sectref{sec:results}, and a summary and discussion are given in \sectref{sec:summary}.

\section{Methods}\label{sec:method}
We follow the method presented in our previous papers \citet{Chen+al:2023} and \citet{Chen:2025} to implement the data-driven bottom boundary and to conduct multi-stage simulations. Here, we describe the particular modification of the method and model setup for this active region. 

\subsection{The Data-Driven Boundary}
The horizontal electric field $\mathbf{E_{\rm h}} = (E_{x}, E_{y})$ that is needed to drive the induction equation is solved from 
\begin{align}
\nabla\times\mathbf{E_{h}} &= -\frac{\Delta B_{z}}{\Delta t} \\
\nabla\cdot\mathbf{E_{h}} &=  -\Omega B_{z},
\end{align}
where $B_{z}$ is the radial component of the vector magnetic field observed by SDO/HMI \citep{HMI,Hoeksema+al:2014}, $\Delta t=720$\,s the cadence of the observations, and $\Omega=1.0\times10^{-5}$\,s$^{-1}$. $\Omega$ represents a rotation applied on a vertical flux tube and is the primary parameter that controls the injection of free magnetic energy into the domain, as described in earlier papers \citep{Cheung+DeRosa:2012,Chen:2025}. We use an educated guess on the basis of the relative motion of the flux concentrations in this active region to determine the value of $\Omega$ as follows.

The evolution of the vertical magnetic field in the photosphere is shown in the top row of \figref{fig:mag}. This active region, as marked in \figref{fig:mag}, comprises two major sunspot pairs, P1--N1 and P2--N2 \footnote{\citet{SunXudong+al:2012} denoted them as P0--N0 and P1--N1.}, and a small parasitic bipole, P3--N3 (N3 merges with N2 and is not marked in the figure), that emerges after February 14, next to the northern edge of the N2 spot. The P3 flux concentration continues to intrude into the area surrounded by the N1, P2, and N2 spots, which causes the destruction of a pre-existing flux rope at the onset of the eruption, as demonstrated later in this paper. The relative shearing and collisional motion between two large sunspot pairs, particularly between N1 and P2, creates a strongly sheared polarity inversion line (PIL) and is suggested to be the key process for generating free magnetic energy in the corona \citep{Toriumi+al:2017b,Chintzoglou+al:2019,Rempel+al:2023}. We estimate the order of magnitude of $\Omega$ from the sunspot motion as follows. The shearing motion is treated as an equivalence of the N1 spot rotating around the P2 spot. In two days, N1 rotated by approximately $\pi/2$ rad with respect to P2, according to a comparison of Panels (b) and (d) in \figref{fig:mag} \citep[see also Figure 3 in ][]{SunXudong+al:2012}, which corresponds to a rotation period of 8 days. In the case of solid body rotation, the vorticity of $1.0\times10^{-5}$\,s$^{-1}$ corresponds to a period of 14.6 days. Control experiments with $\Omega=0$ or $6\times10^{-6}$\,s$^{-1}$ yield no major eruptions by the end of February 15.

\subsection{A Three-Stages Hybrid Strategy}
We modify the two-stage hybrid strategy described in \citet{Chen:2025} to address the combined evolution of the emergence of AR 11158, the formation of coronal plasma structures, and the onset and propagation of the eruption.

\subsubsection{Zero-$\beta$ Model}\label{sec:zb}
The computational domain of a zero-$\beta$ simulation is resolved by $256\times256\times1152$ grid points. The grid spacings are 576\,km and 64\,km in the horizontal and vertical directions, respectively. This setup represents a horizontal domain that is smaller than the actual active region by a factor of 1.875. As in the zero-$\beta$ run in \citet{Chen:2025}, a speed-up factor $f_{\rm sp}=12$ is applied to the time series of the observed magnetic field such that the interval of consecutive magnetograms becomes 60\,s. This option is motivated by the fact that the evolution of the coronal magnetic field is much faster than the photosphere flux emergence and provides a significant reduction in computation time. Moreover, control experiments that employ smaller $f_{\rm sp}$ and the actual active region size yield consistent results in the magnetic energy evolution and the onset of a major eruption, as discussed in \appdref{sec:app_control}. The simulation starts from February 10, 22$^{\rm h}$ UT and is evolved for approximately $6\times10^{5}$ iterations, which covers an equivalent time period of almost 5 days. As shown in \sectref{sec:results}, a strong decay in the magnetic energy is detected at the equivalent time of approximately February 15, 0$^{\rm h}$ UT. The eruption, which involves a consistent evolution of the plasma and magnetic field, needs to be modeled via MHD simulations as described below.

\subsubsection{Radiative MHD Model for the Corona}
The second stage is a radiative MHD run (hereafter, evo. run) that models the coronal plasma before the eruption of interest. The speed-up factor of time is NOT applied in the MHD simulations. The MHD run is started from approximately February 14, 20$^{\rm h}$ UT on the basis of a zero-$\beta$ run snapshot.

A major modification to the method described in \citet{Chen:2025} is that the MHD model represents the one-to-one size of the actual active region. The simulation domain is resolved by a $512\times512\times1920$ mesh, with a horizontal/vertical grid spacing of 540\,km${/}$64\,km. This setup leads to a domain width of 276.48\,Mm and a height of 122.88\,Mm. To construct the initial magnetic field, we calculate a potential field $\mathbf{B}_{\rm P-512}$ from the observed vertical magnetic field, with a periodic side boundary and the assumption of a vanishing field at infinity. The non-potential field of the zero-$\beta$ snapshot $\mathbf{B}_{\rm NP-256}$ can be obtained by deducting the potential field component $\mathbf{B}_{\rm P-256}$ from zero-$\beta$ magnetic field cubes $\mathbf{B}_{\rm z\beta}$. The coordinates of $\mathbf{B}_{\rm NP-256}$ in all three directions are scaled up by a factor of 1.875 to match the size of the actual active region. Then, the cube is interpolated to a $512\times512\times2160$ mesh to fit the grid spacings of the MHD run, and we extract the lower 1920 grid points as $\mathbf{B}_{\rm NP-512}$. Finally, the initial magnetic field is constructed by
\begin{equation}
\mathbf{B}_{\rm init} = \mathbf{B}_{\rm P-512} + \mathbf{B}_{\rm NP-512} + \mathbf{B}_{\rm QS},
\end{equation}
where $\mathbf{B}_{\rm QS}$ is the magnetic field adapted from the part of the domain starting from the photosphere in a fully relaxed quiet Sun simulation similar to that shown in \citet{Chen+al:2022}. More specifically, the lower 1800 grid points of the $\mathbf{B}_{\rm QS}$ cube are extracted from the quiet Sun snapshot, and the upper 120 grid points are filled with a potential field that is calculated from the vertical magnetic field of the top layer.

The initial plasma conditions, including the density, internal energy, and velocities, are also adapted from the quiet Sun simulation in a similar fashion. The thermodynamic variable of upper 120 grid points is filled by a hydrostatic equilibrium and the velocities are set as zero. As we found in simulations for a non-erupting active region \citep{Chen:2025}, the active region corona evolves on the basis of the heating in the coronal volume and is almost independent of the initial plasma conditions.

The MHD model is evolved until the expected eruption of interest occurs, which takes approximately 3.2 hours. We set up a new MHD model to model the dynamics of the eruption. 

\subsubsection{Radiative MHD Model for the Eruption}
The third stage in the hybrid model is an MHD simulation with an extended vertical domain (hereafter, flare run). The flare run has a mesh of $512\times512\times3840$ grid points with the same grid spacing as the evo. run; thus, the domain height is extended to 245.76\,Mm, which provides a large space for the upward eruption.

The flare run is initiated from the snapshot of the evo. run at 300000 iterations, corresponding to February 14, 23$^{\rm h}5^{\rm m}$. The lower 1920 grid points of the initial condition are copied from the evo. run snapshot. The magnetic field in the upper 1920 grid points is a potential field calculated from the vertical magnetic field in the top $x-y$ plane of the evo. run snapshot. The density and internal energy in the upper 1920 grid points are uniform in the horizontal direction and are extrapolated in height following the mean gradient of the uppermost 10\,Mm in the evo. run. This yields a stratification that is very close to hydrostatic equilibrium. The initial velocity in the upper half of the flare run domain is set to zero.

The run cases discussed in this paper, including the production runs described above and control experiments that evaluate the effects of temporal and spatial scaling, are listed in \tabref{tab:list}.

\begin{deluxetable*}{lrcllcc}
\digitalasset
\tablewidth{0pt}
\tablecaption{Summary of simulation cases\label{tab:list}}
\tablehead{
\colhead{Case Name} & \colhead{Grid} & \colhead{Spacing} & \colhead{Time} & \colhead{Time} & \colhead{Speed-up} & \colhead{Volume} \\
\colhead{} & \colhead{$N_{x}{\times}Nz$} & \colhead{$\Delta x$[km]$\times\Delta z$[km]}  & \colhead{started} & \colhead{evolved} & \colhead{factor $f_{\rm sp}$} & \colhead{factor $f_{\rm V}$}
}
\startdata
zero-$\beta$ & $256\times1152$ & $576\times64$ &Feb. 10 22$^{\rm h}$ & $>4.5$ days & 12 & 1.875$^3$ \\
MHD evo.    & $512\times1920$ & $540\times64$ &Feb. 14 20$^{\rm h}$ & $>3.2$ hours & 1 & 1 \\
MHD flare    & $512\times3840$ & $540\times64$  &Feb. 14 23$^{\rm h}$ & $>50$ mins & 1 & 1 \\
\hline
z$\beta$\_$t6$ & $256\times1152$ & $576\times64$  &Feb. 10 22$^{\rm h}$ & $>4.5$ days  & 6 & 1.875$^3$ \\
z$\beta$\_$t3$ & $256\times1152$ & $576\times64$  &Feb. 10 22$^{\rm h}$ & $>4.5$ days  & 3 & 1.875$^3$ \\
z$\beta$\_large & $256\times1080$ & $1080\times128$  &Feb. 10 22$^{\rm h}$ & $>4.5$ days  & 12 & 1 \\
z$\beta$\_large\_$t3$ & $256\times1080$ & $1080\times128$  &Feb. 13 6$^{\rm h}~$ & $>2.5$ days  & 3 & 1 \\
\enddata
\tablecomments{
All the models use $N_{x}=N_{y}$ and $\Delta x=\Delta y$.}
\end{deluxetable*}

\section{Results}\label{sec:results}
\subsection{Emergence of the Active Region and Energy Budget of the Erution}
\begin{figure*}
\includegraphics[width=18cm]{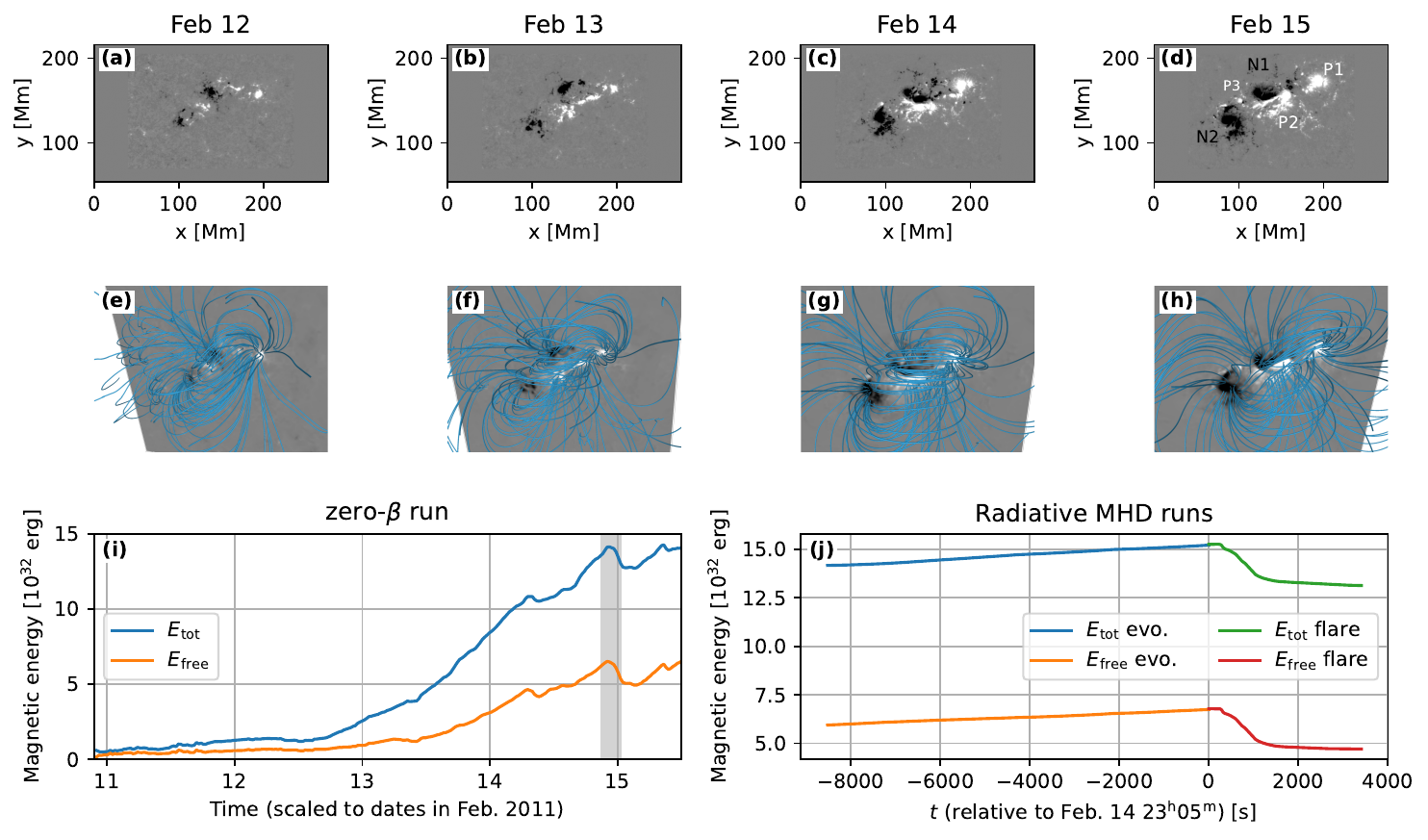}
\caption{The upper panels display the evolution of the observed radial magnetic field of AR 11158. The second row illustrates the coronal magnetic field in the zero-$\beta$ run. Fieldlines are calculated from static seed points that are uniformly distributed in a rectangular area covering the sunspots in the active region. The points of view are chosen according to the location of the real active region on the solar disk. The bottom row shows the evolution of the magnetic energy in the 4 days covered by the zero-$\beta$ run and the rapid decay of the magnetic energy in the radiative MHD runs during the eruption, as indicated by the gray banner.
\label{fig:mag}}
\end{figure*}

The coronal magnetic field in the zero-$\beta$ run across 4 days\footnote{The first three columns show the snapshots that is most close to $0^{\rm h}$ UT of each day, whereas the last column shows the snapshot corresponding to February 14, $20^{\rm h}$ UT, which is identical to the initial state of the MHD evo. simulation.} are displayed in Panels (e) -- (h) of \figref{fig:mag}. The seed points for calculating fieldlines are uniformly distributed in the rectangular area covering the sunspots in this active region. In the latter two days, the coronal fieldlines closely resemble the shape of the observed EUV loops. However, a quantitative investigation of the formation of these loops requires a model with much finer resolution, as suggested in our earlier study on a non-eruptive active region \citep{Chen:2025}.

A crucial parameter of flare-productive active regions is the free magnetic energy stored in the corona. Evolution of the volume-integrated total magnetic energy $E_{\rm tot}$, as given by
\begin{equation}
E_{\rm tot} = \int \frac{1}{2\mu_0}\mathrm{B}^{2}\,dV,
\end{equation} 
and free magnetic energy $E_{\rm free}$, which is estimated by
\begin{equation}
E_{\rm free} = E_{\rm tot} - \int \frac{1}{2\mu_0}\mathrm{B^{2}_{\rm P}}\,dV,
\end{equation}
in the zero-$\beta$ run is plotted in Panel (i) of \figref{fig:mag}\footnote{On the basis of the setup of the zero-$\beta$ run, the magnetic energy and time are multiplied by a factor of $f_{\rm V}=1.875^{3}$ and $f_{\rm sp}=12$, respectively, such that they can be directly compared with quantities of the MHD simulations}. $\mathbf{B}_{\rm P}$ stands for the potential field calculated from the vertical magnetic field at the bottom of the domain. The rapid increase in magnetic energy starts from February 13, similar to the trend found in observations \citep{SunXudong+al:2012}. In our model, the magnetic energy increases to approximately $1.4\times10^{33}$\,erg with a very steady rate until a time corresponding to February 14, $23^{\rm h}$ UT; meanwhile, the free magnetic energy accumulates to approximately $6.5\times10^{32}$\,erg. We note that this energy includes a large contribution from the curved/twist magnetic field outside the core region of the domain, which is an effect of the constant $\Omega$ and the periodic side boundary. An abrupt decay in the free magnetic energy, as indicated by the gray mask in Panel (i), indicates a major eruption, which occurs approximately 3 hours earlier than the actual X class flare on real Sun (February 15 $2^{\rm h}$ UT). Given the 100-hour evolution time since February 10, 22$^{\rm h}$ UT, this difference corresponds to an error of 3\%.

The evo. and flare runs describe the course of the eruption from development to onset. \figref{fig:mag}(j) shows the joint evolution of the total and free magnetic energy since 2.5 hours before the flare onset. The time $t$ hereafter is relative to the beginning of the MHD flare run. The free magnetic energy steadily increases and starts to decrease at $t=219$\,s. The fast decay lasts for approximately 1000\,s, during which $1.94\times10^{32}$\,erg magnetic energy is released.
  
\subsection{Magnetic Structure and Plasma Dynamics of the Eruption}
\begin{figure*}
\includegraphics[width=18cm]{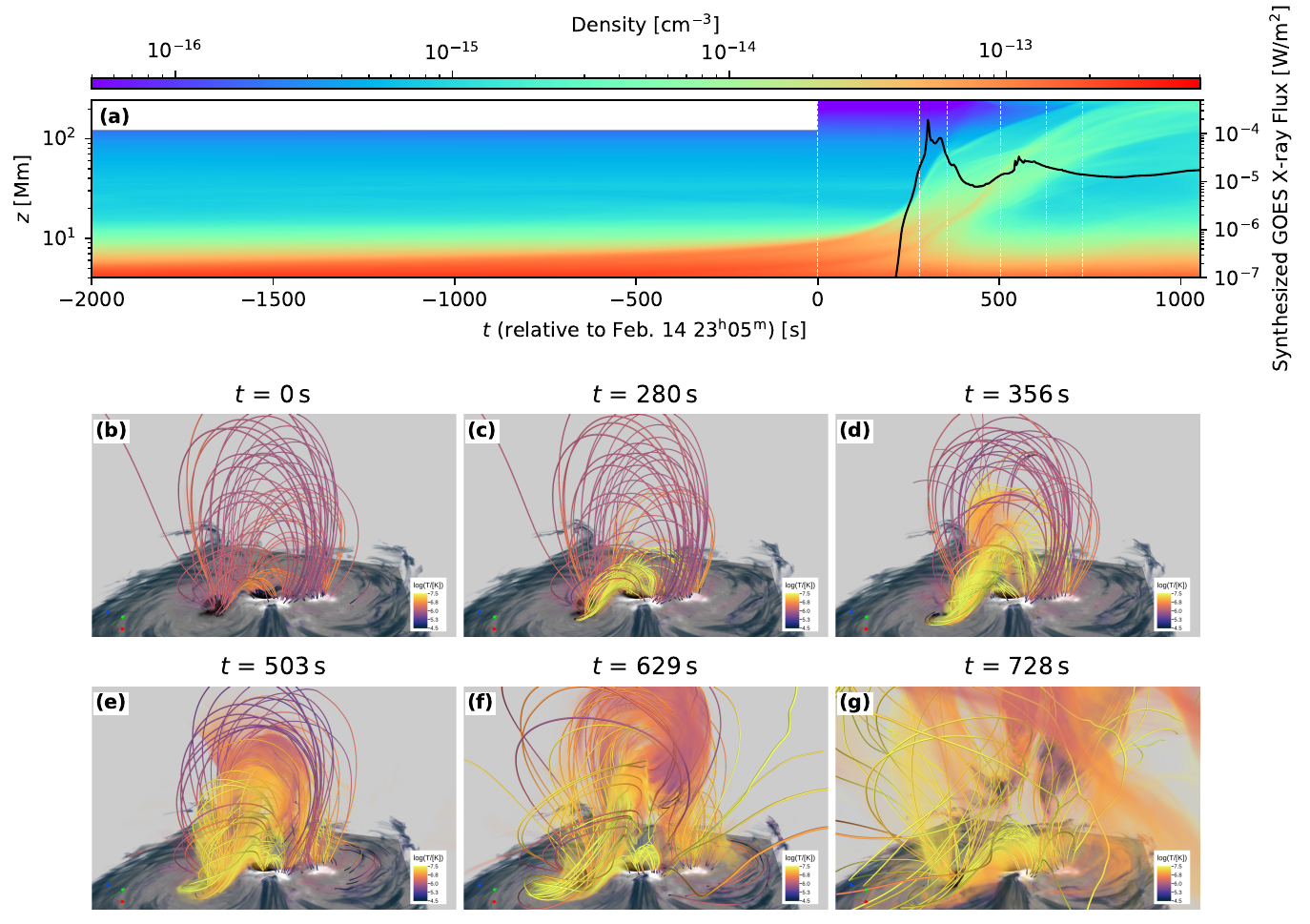}
\caption{Course of the eruption. Panel (a) shows the horizontally averaged density from the combined data of the evo. and flare runs, which illustrate the slow rise and eruption of a plasma-hosting magnetic flux rope. The overlay black line plots the synthesized GOES 1-8\AA~soft X-ray flux, which indicates a flare above X class. Panels (b)--(g) present 3D visualizations of the plasma and magnetic field structures of the eruption. The corresponding time stamps are marked by the white dashed lines in Panel (a). The opacity of the features is chosen on the basis of the plasma density in the erupted flux rope, whereas lower and higher values are made transparent. The magnetic field lines are calculated from seed points that are randomly distributed in the coronal volume above the sunspots. The color of the density features and fieldlines reflects the plasma temperature on a logarithmic scale.
\label{fig:vapor}}
\end{figure*}

We present in this section the eruption of a flux rope and the large-scale dynamics in the model atmosphere. 

\subsubsection{Slow Rise and Fast Eruption of the Flux Rope}
We use density structures to trace the kinematic evolution of the flux rope, as in our previous studies \citep{WangCan+al:2023} and in observations where ejected plasma is tracked \citep[][ and reference therein]{ChengXin+al:2020}. \figref{fig:vapor}(a) presents the horizontally averaged plasma density as a function of time. The 3000\,s time period covers the slow rise of the flux rope before the rapid eruption. The height $z$ starting from 4\,Mm is displayed on a logarithmic scale to highlight the lower part of the domain.

We calculate the soft X-ray flux with a temperature response function of the GOES-15 satellite, which is overlaid on the density profile. The peak flux of $1.97\times10^{-4}$\,W m$^{-2}$ is found at $t\approx300$\,s, which corresponds to an X2 class flare and is close to the X2.2 class of the real flare \citep{Schrijver+al:2011}. The similarity in the soft X-ray flux is very likely only a {\em coincidence} because real flares undergo physical processes, e.g., those related to nonthermal particles \citep{Aschwanden+al:2016_III,Warmuth+Mann:2016b}, that are beyond the capability of the current numerical model. Therefore, the actual value of the soft X-ray flux is only compared with other simulations under similar approximations and is considered an indicator of the strength of energy release and plasma heating. This flare is thus far the largest event, compared with early MURaM simulations in the literature that produced flares corresponding to lower M \citep{Chen+al:2023L,Rempel+al:2023} or C class \citep{Cheung+al:2019}, with magnetic energy release on the order of $10^{31}$\,erg.

For more than 1000\,s before the eruption, the flux rope steadily rises at a very small speed of approximately 2\,km/s and progresses to rapid eruption around $t=250$\,s during the impulsive phase of the flare. The lead edge in the density profile is a piston-driven shock (presented later with more details in \sectref{sec:shock}). The main body of the flux rope rises at an average speed on the order of 200\,km/s. Although the domain height is the maximum that our computational resources can afford, the shock and ejected plasma reaches the top boundary at approximately $t=420$\,s and $t=700$\,s, respectively. The top boundary is open for outflows but may still affect the upward propagation of the CME. Observations \citep[e.g., Figure 9 in ][]{Schrijver+al:2011} revealed a radial expansion front between 40--100\,Mm above the solar surface with a deduce speed of 300\,km/s at 4\,min after the onset, which is similar to the propagation in this simulation.

\subsubsection{Large-scale Magnetic Structure in the Eruption}
We briefly present the magnetic and plasma structures in the early stage of the eruption in Panels (b) -- (g) of \figref{fig:vapor}. Before eruption, the flux rope can be found along the PIL between N1 and P2 and is accompanied by sheared arcades on both sides. Cool plasma is hosted in the dip formed by a twisted magnetic field. Panel (c) shows the snapshot in the impulsive phase of the flare, during which reconnections occur between the pre-existing flux rope with ambient arcades and gives rise to 10\,MK plasma in the core area between N1 and P2, which is the location of the observed X-ray sources \citet{Zharkov+al:2011,WangShuo+al:2012}. Panel (d) displays a snapshot shortly after the flare peak. The two hot arcades connecting from the inner spots N1 and P2 to the outer spots P1 and N2, respectively, now reconnect to form a long sigmoid, which is considered as a new large flux rope that roots in P1 and N2, and at the same moment, post-flare loops connecting N2 and P1 are formed below the reconnection site.

The snapshots in the gradual phase are presented in Panels (e) -- (g) of \figref{fig:vapor}. The long sigmoid continues to rise and push the plasma upward with ongoing reconnection below, which is similar to the mechanism found in more generalized numerical experiments \citep{JiangChaowei+al:2021}. New post-flare loops continue to form left (solar east or lower $x$) to right (solar west or higher $x$). The propagation of reconnection is supposed to drive the extension of flare ribbons, which is discussed later in this paper. Panel (g) illustrates the reconnection between the envelope field on the two sides of the erupted flux rope and the formation of longer and higher post-flare loops that are filled with plasma over 10\,MK.

\subsubsection{Plasma Dynamics in the Large-scale Corona}
\begin{figure*}
\includegraphics[width=18cm]{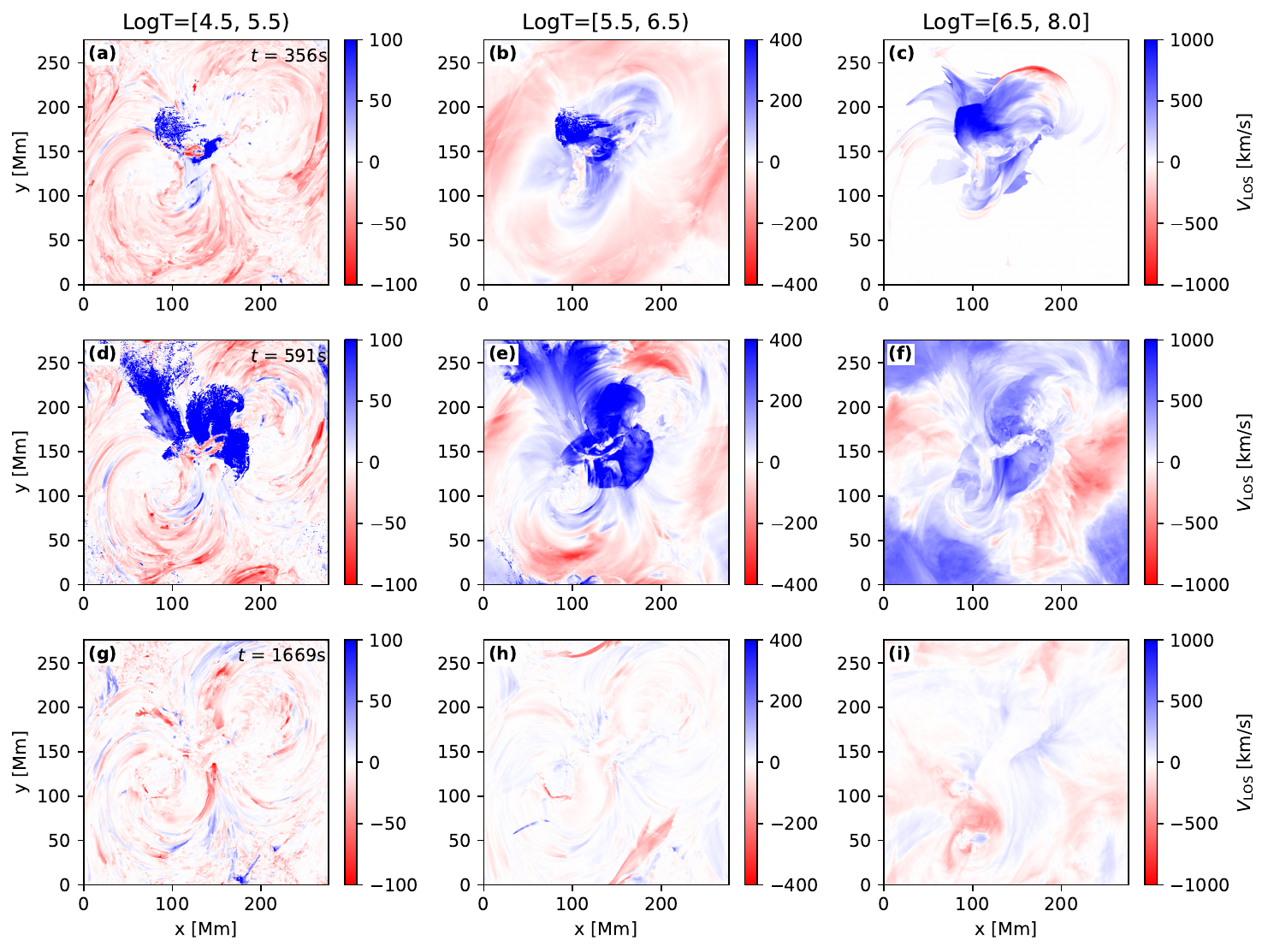}
\caption{The emission measured line-of-sight velocity from a top view. It represents the Doppler velocity from a spectroscopic observation of an active region near the solar disk center. The original outputs of the emission measures and line-of-sight velocities at an interval of $\log T=$0.1 are binned to three temperature ranges, as indicated at the top panel of each column, and the rows correspond to three snapshots as marked in the first panel of each row.
\label{fig:doppler}}
\end{figure*}

We analyze the emission measured weighted velocity $v_{\rm LOS}$, which represents the Doppler velocity of a spectral line forming in the given temperature range, as in our previous studies \citep{Chen+al:2023, Chen:2025}. 
$v_{\rm LOS}$ within a given temperature range $[T_{\rm L}, T_{\rm H}]$ is calculated as
\begin{equation}
v_{\rm LOS} = \frac{\sum\limits_{T_{i}=T_{\rm L} }^{T_{\rm H}} v_{z}(\log T_{i}){\rm EM}_{z}(\log T_{i})}{\sum\limits_{T_{i}=T_{\rm L} }^{T_{\rm H}} {\rm EM}_{z}(\log T_{i})}.
\end{equation}
${\rm EM}_{z}(\log T_{i})$ is a standard MURaM output for analysis, as described in \citet{Chen+al:2022}, which records the emission measures integrated along a given axis (the vertical axis $z$ here) for temperatures from $\log T$=4.5 to the highest value in the domain with an interval of $\Delta \log T = 0.1$. Similarly, $v_{z}(\log T_{i})$ is the line-of-sight average velocity for the same temperature discretization of ${\rm EM}_{z}$. Moreover, this average uses the emission measure as a weight, when the contribution of each cell in a vertical column is counted into the corresponding temperature interval. \figref{fig:doppler} presents $v_{\rm LOS}$ in three temperature ranges that capture the chromosphere and transition region plasma, warm coronal plasma of million K, and hottest plasma around and above 10\,MK.

The first row of \figref{fig:doppler} shows a snapshot shortly after the peak of the flare. The low-temperature Doppler map reveals a strong downflow up to 100\,km/s at the left end of the PIL between N2 and P1. The two ribbons of downflow outline the footpoints of the post-flare loops formed beneath the reconnection site, as illustrated in \figref{fig:vapor} (d). The upflows embracing the downflow indicate fast ejection of chromosphere plasma. The upflow pattern, particularly the one above P3 in the intruding bipole, can be observed at all temperatures with increasing speed to over 1000\,km/s in the hottest bins.

The second row displays Doppler maps in the gradual phase of the flare. The downflow ribbons significantly extend along the PIL, as we expected from the development of magnetic reconnection, as shown in \figref{fig:doppler}(e) and (f). Moreover, the eruption produces more spatially extensive upflows of a few hundred km/s. \citet{Fan+al:2024} reported a significant fieldline expansion at similar locations, which may explain the dimming observed during this eruption \citep{Dissauer+al:2018}. 

The bottom row presents a snapshot when the atmosphere is recovering from the highly dynamic state caused by the eruption. The time scale might not be realistic because the real eruption impacts a much larger volume than the domain size of the simulation and lasts significantly longer. Nevertheless, during this stage, the plasma velocities at all temperatures decrease to their typical values during a quiescent period.

Last but not least, the upward eruption of the magnetic structure during the impulsive and peak phases of the flare is accompanied by strong horizontal expansion as well, which eventually leads to a downward push at the outer rim of the active region, as shown by the increasing and sharpening downflow rims in Panels (b) and (e) of \figref{fig:doppler}. This gives rise to intriguing interactions between different layers of the atmosphere, as presented in detail in the following section.

\subsection{CME Driven Shock and the Chromospheric Moreton Wave}\label{sec:shock}
\begin{figure*}
\includegraphics[width=18cm]{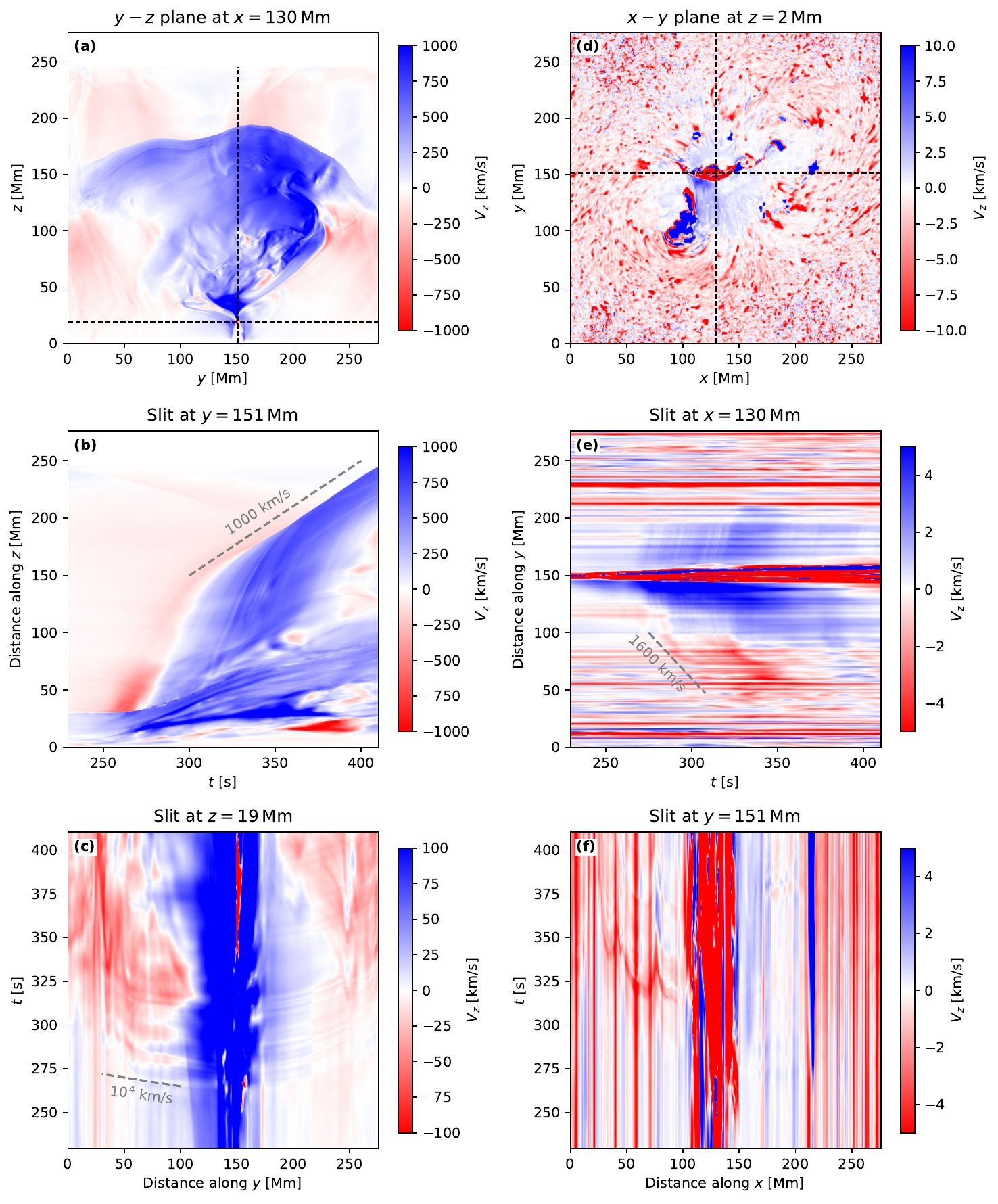}
\caption{Piston-driven shock in the corona and the Moreton waves in the chromosphere. In the left column, Panel (a) presents the vertical velocity $v_{z}$ at $t=356$\,s in a $y-z$ plane placed at $x=130$\,Mm; Panels (b) and (c) are time-distance diagrams of $v_{z}$ along the vertical and horizontal slits, which are placed at the dashed lines in Panel (a), respectively. In the right column, Panel (d) shows $v_{z}$ in a horizontal layer at 2\,Mm height; Panels (e) and (f) are time-distance diagrams of $v_{z}$ along the $y$- and $x$-slits, as indicated by the dashed lines in Panel (d), respectively.
\label{fig:shock}}
\end{figure*}

In \figref{fig:shock} (a), we present the vertical velocity $v_{z}$ in a $y-z$ plane that cuts through the core region of the eruption near the peak of the flare, which is the same time as \figref{fig:vapor} (d) and the top row of \figref{fig:doppler}. The plasma after the piston-driven shock has an upward velocity of 500\,km/s, in contrast to the mild 200\,km/s downward velocity in the background. In comparison, \citet{Afanasyev+al:2023} and \citet{Fan+al:2024} reported velocities above or close to 1000\,km/s after the shock front. The highest speed in this vertical slice is found immediately above the reconnection site at $z\approx20$\,Mm, which is up to 1500\,km/s and accompanied by a 500\,km/s downflow right below, which appears as a red point in \figref{fig:shock} (a).

Panels (b) and (c) display two time-distance diagrams of $v_{z}$ along a vertical and a horizontal slit that are placed at locations indicated by the two dashed lines in \figref{fig:shock} (a), respectively. Panel (b) shows that the upward-propagating disturbance is set off in the impulsive phase at approximately $t=270$\,s. The velocity gradient increases and develops to a shock (when it is resolved by 5 grid points in the vertical direction) at $t\approx320$\,s and $z\approx170$\,Mm. Afterward, the shock propagates at a rather constant speed of approximately 1000\,km/s. Moreover, \figref{fig:shock} (b) shows similar branches, as illustrated in \figref{fig:vapor} (a), where we see a fast propagating shock front and a flux rope that rise at a small speed.

The diagram in \figref{fig:shock} (c) aims to capture the impact of the eruption at the outer part of the active region. We find numerous thin ridges that extend from the spine, which is the upward erupting flux rope, to both sides. The slopes of the ridges indicate a propagation speed on the order of 10$^{4}$\,km/s. After $t=275$\,s, a few ridges (on the lower $y$ side) exhibit downflows, which means that the atmosphere above may exert a downward push to that below this layer and  even deeper. 

Similar behaviors were also reported by \citet{Afanasyev+al:2023} and \citet{Fan+al:2024}. Unfortunately, their model may not have a setup that allows for studying the dynamics in the lower atmosphere. Therefore, we examine this effect in our simulation in a deep horizontal layer at 2\,Mm height in the chromosphere, as shown in \figref{fig:shock} (d). In the outer rim of the active region, which is dominated by downflows, a belt of slightly increased downward velocities (e.g, at $y=60$\,Mm) can be discerned. This disturbance is better revealed in Panel (e) by a time-distance diagram along the $y$-slit placed at the exact location of the $y-z$ plane. After $t=275$\,s, we see in this deep chromosphere layer a propagating downflow wave front to the south, which will generate a signal on the red wing of chromospheric spectral lines such as how Moreton waves were discovered \citep{Moreton+Ramsey:1960}. The propagation speed of approximately 1600\,km/s is surprisingly large compared with the typical sound speed in the chromosphere. It follows the long-known argument that such chromospheric waves cannot be sustained in the chromosphere but rather imprint the impact of waves/shocks propagating in the corona \citep{Uchida:1968}.

Panel (f) of \figref{fig:shock} displays a time-distance diagram along the $x$-slit marked by the horizontal dashed line in Panel (d). The imprint of coronal waves/shocks is only found east of the active region, and together with \figref{fig:shock} (e), it clearly demonstrates the asymmetric propagation of the Moreton wave.

\subsection{Formation of Flare Ribbons}

\subsubsection{Plasma Dynamics in Flare Ribbons}
\begin{figure*}
\includegraphics[width=18cm]{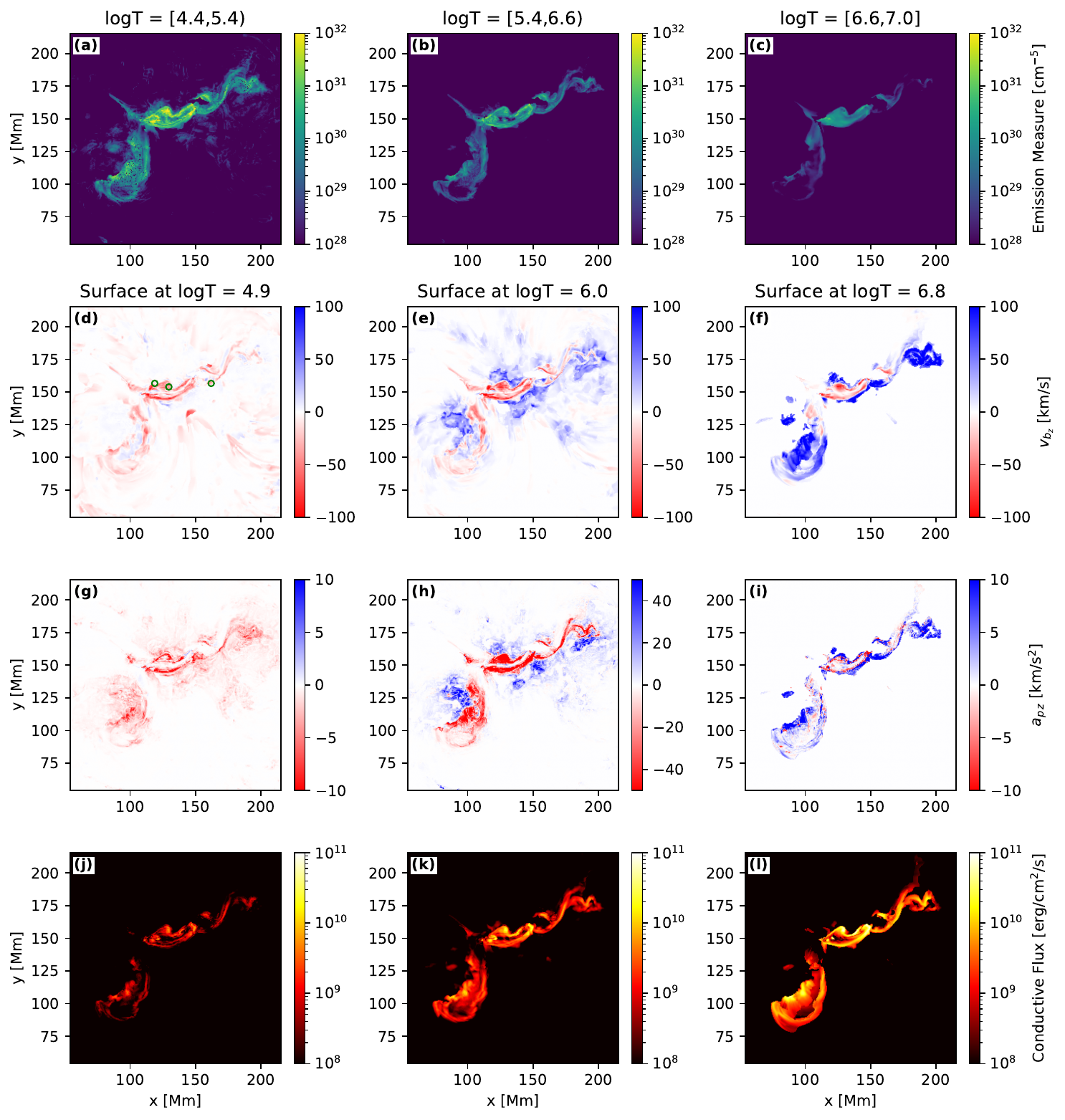}
\caption{The flare ribbons at $t=410$\,s. Only the lowermost 12.8\,Mm (200 grid points) are considered in this analysis to highlight the emission in the lowest part of the domain and to exclude the contribution from ejected chromospheric plasma with the erupted flux rope. The first row displays emission measures binned to three wide temperature ranges. The second row shows the vertical component of field-aligned velocities, i.e., the upward evaporation flows, on the surface at a certain temperature, as indicated above Panels (d)--(f). Panels (g)--(i) present the vertical acceleration driven by the pressure gradient force along the magnetic field. Panels (j)--(l) illustrate the thermal conduction flux on these three temperature surfaces.
\label{fig:ribbon}}
\end{figure*}

Bright ribbons are the most prominent observational manifestations of solar flares in the lower atmosphere. The first row of \figref{fig:ribbon} displays the emission measures during the gradual phase in the domain below 12.8\,Mm, where the contribution is mostly from the heated chromosphere and the lower legs of the post-flare loops. The emission measures over a broad temperature range are integrated into three bins. The brightest ribbons are found at the lower temperatures along the PIL. The flare ribbons at this moment have extended to the outer sunspots, N1 and P2, which are the footpoints of the large flux rope that is formed through reconnections during the eruption. The locations are consistent with the magnetic configuration shown in \figref{fig:vapor}.

Instead of the line-of-sight integrated velocity shown in \figref{fig:doppler}, we illustrate the plasma dynamics in flare ribbons via the velocity on a constant temperature surface. Such surfaces are naturally 3D structures in space. We use a linear interpolation on $\log T$ to obtain the height $z$ of a constant temperature surface as an $x-y$ map and interpolate column-wise from 3D cubes of quantities of interest to extract the values on this temperature isosurface.

Panels (d)--(f) reveal chromospheric evaporations and condensations in the flaring atmosphere in three layers through the field-aligned vertical velocity $v_{\rm b_{z}}$, which is given by 
\begin{equation}
v_{\rm b_{z}} = \left(\mathbf{v}\cdot\mathbf{b}\right)b_{z},
\end{equation}
where $\mathbf{b} = \mathbf{B}/|\mathbf{B}|$ is the unit vector of the magnetic field. Moreover, the pressure gradient is the major driving force of field-aligned dynamics in response to energy deposition in the chromosphere. We present the pressure-driven vertical acceleration $a_{p\,z}$, which is evaluated by 
\begin{equation}
a_{p\,z} = \frac{(\nabla p\cdot\mathbf{b})\,b_{z}}{\rho},
\end{equation}
on the same temperature surfaces in Panels (g) -- (i).

Strong downflows are found in temperatures over one MK at locations where the brightest ribbons are observed in the center of the active region and are cospatial with the strongest downward pressure gradient acceleration. Slightly weaker condensations are present in the extended ribbons. On the hottest surface, which corresponds to a slightly higher geometric height, the partition of upflows prevails, whereas some points along the ribbon may still exhibit downflows. An extensive area off the flare ribbons is dominated by upflows from one MK. This is also consistent with the blue patches in \figref{fig:ribbon} (h), where the pressure gradient drives very weak condensations at lower temperatures but evident evaporations at higher temperatures.

The dynamics observed in the simulation are attributed to energy deposition in the chromosphere. In this simulation, conduction is the main agent that transports the magnetic energy that is released by reconnections in the corona down to the lower atmosphere. The bottom row of \figref{fig:ribbon} displays the thermal conduction flux in the three layers with increasing temperature (increasing height as well). A downward energy flux greater than $10^{11}$\,erg/cm$^{2}$/s is present in the higher layers, which is comparable to the strongest energy injection rate assumed in radiative hydrodynamics calculations for solar flares \citep[e.g., ][]{Allred+al:2015,Reep+al:2015,Kowalski+al:2017}, whereas the energy flux decays significantly with height and becomes the order of $10^9$\,erg/cm$^{2}$/s on the surface at lower temperatures. This decay indicates a deposition of energy in this volume, particularly along the flare ribbons, where significant condensations and evaporations are driven by downward/upward pressure gradient forces. 

We mark the locations of the sun-quake sources found by \citet{Kosovichev:2011} and \citet{Zharkov+al:2011} during this flare in Panels (d) and (k). The observed sun-quake sources, which should be in a further deeper layer, are more or less cospatial with points of strong condensations, i.e., a high energy injection rate. Moreover, two of them appear to be the footpoints of the pre-existing flux rope along the central PIL, and the middle one is right beneath the reconnection site in the early stage of the flare, as illustrated in \figref{fig:vapor}. Although the present model does not account for all the physics needed to model the trigger of the phenomenon, the comparison suggests a nontrivial similarity between the energy deposition site in the simulation and that in the real flare.

\subsubsection{Evolution of the flare ribbons and Coronal Magnetic Reconnection}

\begin{figure*}
\includegraphics[width=18cm]{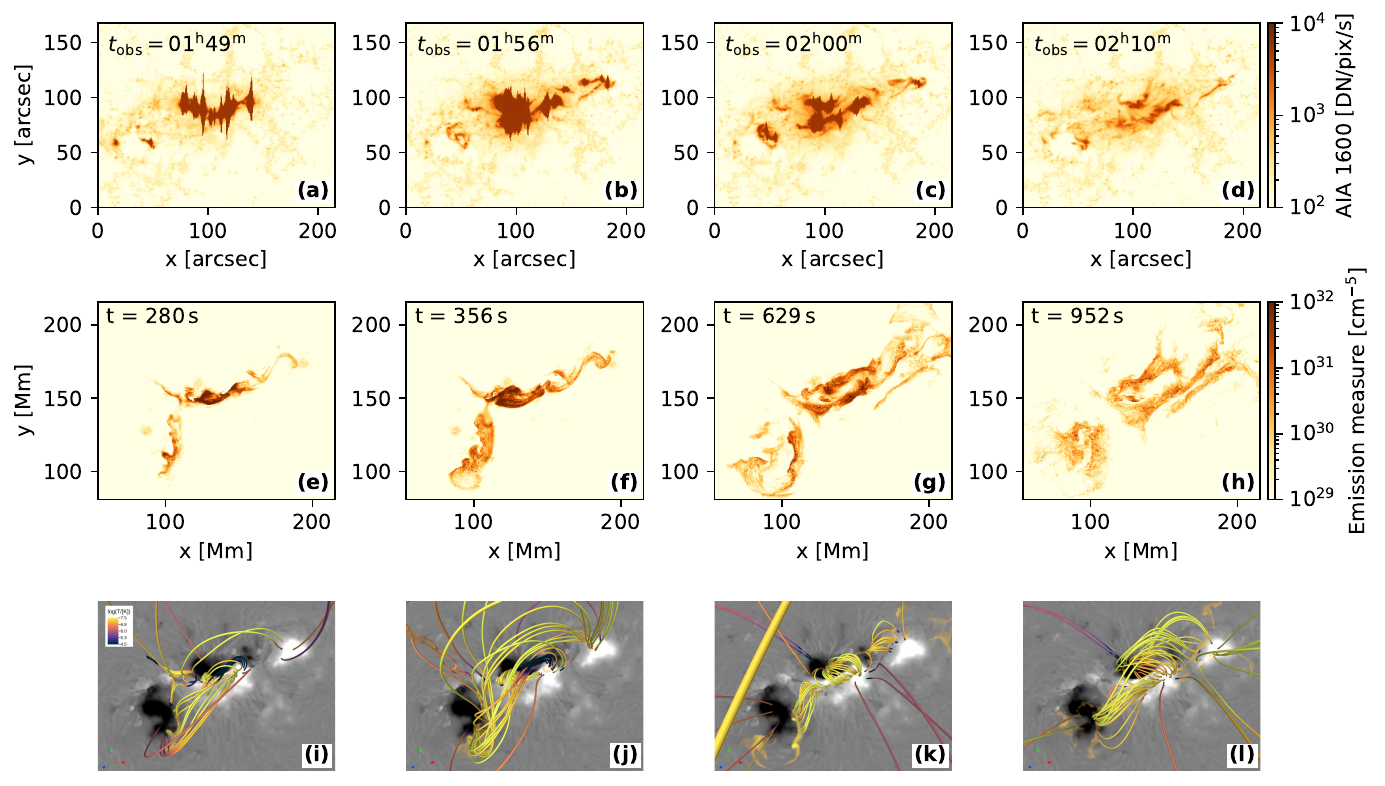}
\caption{Comparison of the flare ribbons in the observation and simulation. The top row displays the evolution of flare ribbons captured by the AIA 1600 channel. The middle row shows the flare ribbon in the simulation, which is manifested by the emission measure in the lowermost 200 grid points at all temperatures. The bottom row presents magnetic field lines that are closely related to flare ribbons. The grayscale images are $B_{z}$ in the photosphere of the simulation, overlaid by a horizontal slice at 2.5\,Mm height showing plasma temperatures, as indicated by the color table. Only temperatures higher than 10\,MK are displayed, whereas lower values are transparent. Two groups of magnetic field lines are displayed: one group is calculated from seed points that are randomly distributed in regions with bright ribbons with a bias to high density and the other group has a bias to high temperature. The color of the fieldlines represents the plasma temperature, as indicated by the same color table of the slice.
\label{fig:ribbon_mag}}
\end{figure*}

Finally, it is intriguing to examine the extent to which the flare ribbons in the model may resemble the observation of this flare and the magnetic configuration that directly causes the evolution of the ribbon. The AIA 1600\AA~channel images covering the change in the flare ribbons from the impulsive to the gradual phases are presented in the top row of \figref{fig:ribbon_mag}. We note that the structures along the y-axis in Panels (a) and (b) are caused by saturation. The observation is compared with the total emission measure at all temperatures in the lower atmosphere, as shown in \figref{fig:ribbon_mag} (e)--(h). The bottom row presents a top-down view of the magnetic field in a similar field of view.

In the impulsive phase shown in the first column in \figref{fig:ribbon_mag}, both the observation and simulation show bright ribbons along the central PIL and a brightening on the right side of the N2 spot. During the peak, while the brightest emission remains in the active region core, the ribbons extend to the center of the P1 spot, with bright patches above strong flux concentrations. The gradual phase in the third column demonstrates that the ribbons in both the observation and model start to fade and exhibit separating motion in the N--S direction. Finally, as time progresses, the two main ribbons clearly become more separated with significantly reduced brightness.

The magnetic field configuration reveals a crucial process in the early stage of the flare between the P3 flux concentration and the pre-existing flux rope along the PIL. As illustrated in Panel (i), part of the flux rope rooted at N1 is involved in a reconnection with the intruding P3 and is redirected to N2, resulting in the brightening observed on the right side of the N2 spot. 

At the moment shown in Panel (j), reconnection between the larger arcades creates long hot loops rooted at the conjugate brightening in the two outer sunspots P1 and N2. Thus, the E--W extension of the ribbon is a manifestation of the continuous reconnections in the corona.

The latter two snapshots presented in Panels (k) and (l) illustrate, in closer view than those presented in \figref{fig:vapor}, the formation of a corridor of post-flare loops. As the large flux rope rises and triggers reconnection between the large scale envelope field, as shown in \figref{fig:vapor} (g), longer and higher loops with apparently weak shear form, whose footpoints map the expanded flare ribbon seen in the emission features.

\section{Summary and Discussion}\label{sec:summary}
In this paper, we present the application of the data-driven branch of the MURaM code to a well-known flare-productive active region. The key results are summarized as follows. 

The complete three-stage model demonstrates the emergence of the magnetic field from the photosphere to the corona and a major solar eruption that releases $10^{32}$\,erg magnetic energy and yields a soft X-ray flux on a similar order of magnitude as the real event. In the large-scale atmosphere, the eruption drives bulk mass ejections and a dome-like shock front; meanwhile, the coronal eruption delivers a sufficiently strong impact on the outer rim of the active region, which penetrates to the chromosphere and creates a fast-propagating Moreton wave. The released magnetic energy is deposited in the chromosphere by strong downward conduction fluxes that are greater than $10^{11}$\,erg/cm$^2$/s. Local pressure peaks drive chromospheric condensation and evaporation. The temporal evolution of the bright flare ribbons in the simulation, which maps the development of reconnections in the coronal magnetic field, is consistent with the observation of this flare.

Observations have suggested that the seminal structure of this eruption was a flux rope along the highly sheared PIL between the N2 and P1 spots. Despite the different magnetic initial conditions and boundary driving methods, data-driven simulations of AR 11158 have consistently reproduced such a magnetic structure \citep{Inoue+al:2014,Hayashi+al:2018,Afanasyev+al:2023,Fan+al:2024}. In our model, we use a simple approach of \citet{Cheung+DeRosa:2012}. Although the outer part of the active region exhibits some artificial twist at the large scale due to the constant $\Omega$ and horizontal size boundary, the choice of $\Omega=1.0\times10^{-1}$\,s$^{-1}$ is able to build an appropriate flux rope that creates an eruption of a similar strength as the real flare.

The plasma dynamics in flare loops have been thoroughly investigated via classical hydrodynamics calculations \citep{Fisher+al:1985}. Idealized 3D MHD models suggest that the locations of flare ribbons reflect the cross sections of magnetic separatrices in the lower atmosphere \citep[e.g., ][]{Janvier+al:2014,Savcheva+al:2015}. The ingredients of these models must be combined in 3D radiative MHD simulations to yield flare ribbons that can be compared with observations, as performed by \citet{Cheung+al:2019} in an MURaM simulation inspired by observations. 

Furthermore, reproducing the observed ribbons in this particular flare is more challenging. \citet{Inoue+al:2014} used magnetic properties to project the location of flare ribbons in the model, which appeared to be similar to the observed ribbons. The models presented by \citet{Afanasyev+al:2023} and \citet{Fan+al:2024} stand for the state of the art of data-driven models of this active region thus far in the literature. The former employed a simplified plasma energy equation but was able to illustrate the location of flare ribbons via a temperature proxy. This proxy is different from observed chromospheric flare ribbons because, in that model, the lower boundary actually represents the coronal base, and temperature evolution is affected by a boundary of fixed values. The latter improved the energy equation and reproduced coronal EUV observations; however, unfortunately, the bottom boundary has a fixed chromospheric temperature and density, which still limits investigations of flare ribbons and the chromospheric impact of the coronal shock. In comparison, our model has a stratified solar atmosphere starting from the photosphere, which is consistent with the layer of the observed magnetic field, and the chromospheric density and temperature dynamically evolve in response to the energy injection from the corona. Therefore, to the best of our knowledge, the work presented in this paper may be the first model that self-consistently describes the evolution of flare ribbons, which resembles a real flare in a one-to-one fashion.

In addition to studying this particular active region, which has been performed rather thoroughly, the simulations presented in this paper could also be useful for investigating, in general, the gradual evolution of a flux rope before eruption and the trigger of onset. The data-driven MURaM code can be of cause applied in general to other observed active regions, as long as a time series of the vertical magnetic field is available. It is also interesting to test in the further how our simulations perform when they are driven by the electric field used in other models \citet{Kazachenko+al:2014,Fisher+al:2020} or a more sophisticated characteristic boundary developed by \citet{Tarr+al:2024}.

\begin{acknowledgments}
F.C. is supported by National Science Foundation of China No. 12422308 and No. 12373054, and by the National Key R\&D Program of China under grant 2021YFA1600504. This work benefits from discussions during the ISSI workgroup ``Data-driven 3D Modeling of Evolving and Eruptive Solar Active Region Coronae". The visualizations shown in \figsref{fig:mag}, \ref{fig:vapor}, and \ref{fig:ribbon_mag} are created by VAPOR \citep{vapor}.
\end{acknowledgments}

\begin{contribution}
F.C. conducted the numerical simulations, performed the analysis and visualization, and wrote the texts.
\end{contribution}

\appendix

\section{Control experiments on the temporal and spatial scaling factors in zero-$\beta$ models}\label{sec:app_control}
\begin{figure}
\includegraphics[width=8.5cm]{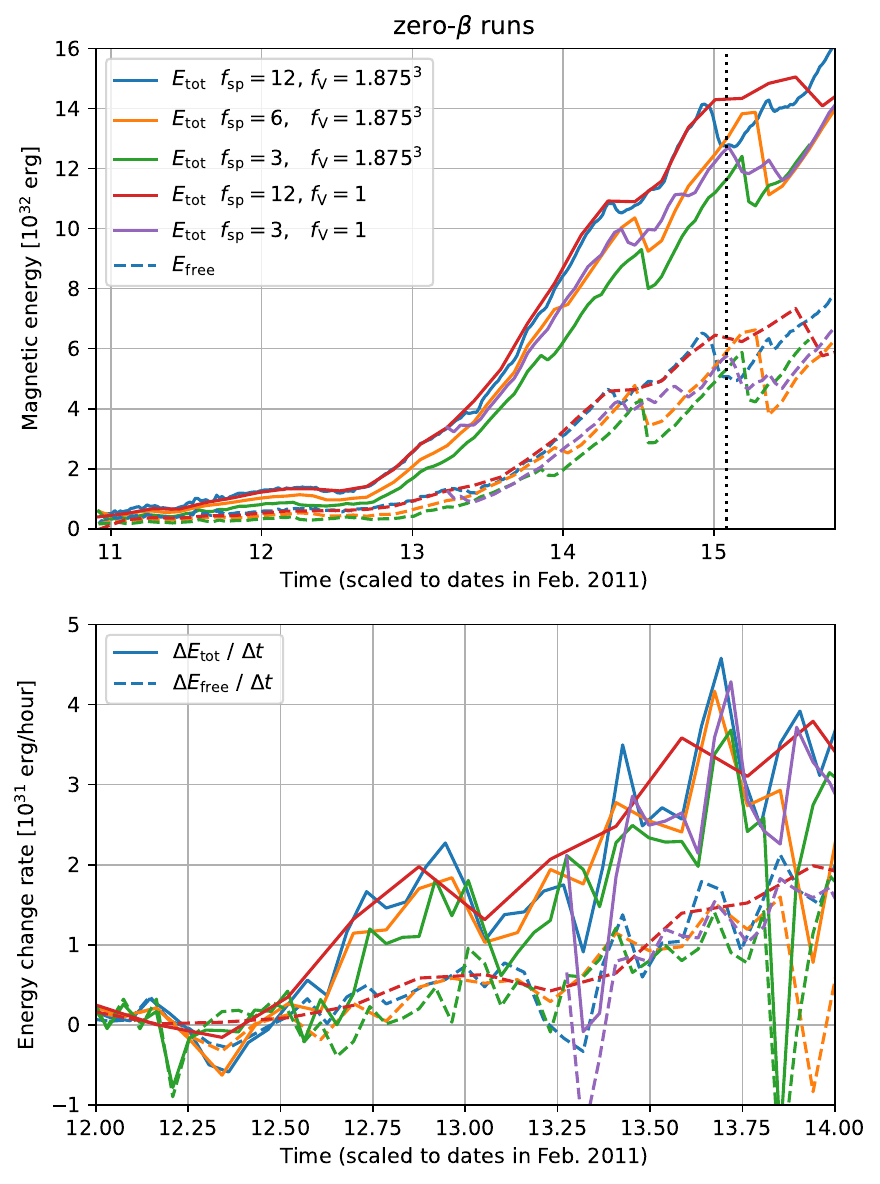}
\caption{Comparison of magnetic energy in zero-$\beta$ runs with different speed-up and volume scaling factors. The upper panel presents the evolution of the total ($E_{\rm tot}$, solid lines) and free ($E_{\rm free}$, dashed lines) magnetic energy over time. The lower panel shows the energy change rate, i.e., the time derivative of the magnetic energy between February 12 and February 14, following the line colors noted in the upper panel. Four cases are compared. The time and energy of each case are multiplied by its corresponding speed-up factor $f_{\rm sp}$ and the volume scaling factor $f_{\rm V}$, respectively. The vertical dotted line indicates the time of the real X2.2 flare.
\label{fig:emag_scaling}}
\end{figure}

\begin{figure}
\includegraphics[width=8.5cm]{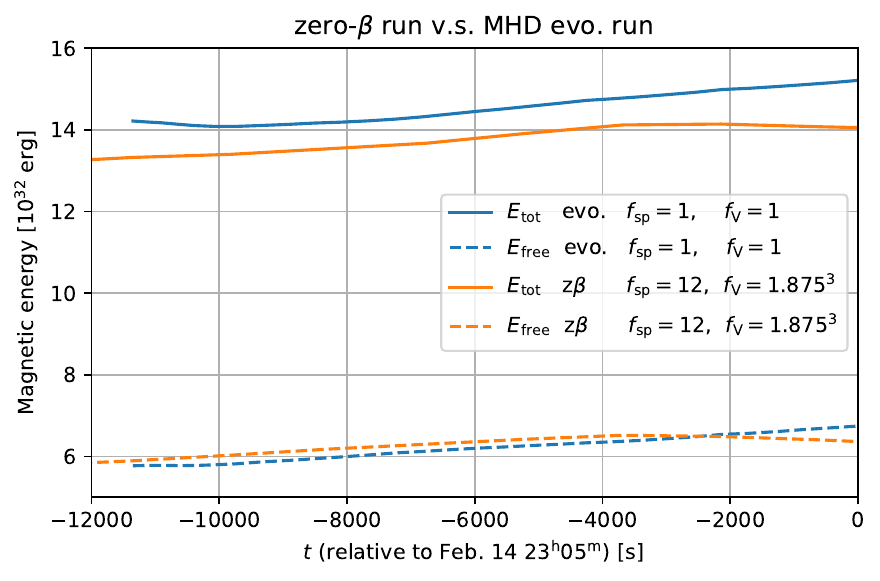}
\caption{Comparison of the magnetic energy in the zero-$\beta$ and the MHD evo. runs shown in \figref{fig:mag}. The time and energy of the zero-$\beta$ run are multiplied by $f_{\rm sp}$ and $f_{\rm V}$, respectively.
\label{fig:emag_zbevo}}
\end{figure}

As described in \sectref{sec:method}, the evolution of the observed magnetic field that drives the zero-$\beta$ model is accelerated by a factor $f_{\rm sp}$, and the domain of the zero-$\beta$ model represents a smaller region than the actual active region. The robustness of the results is substantiated by further control experiments.

The z$\beta$\_$t6$ run employs the same domain setup as the reference zero-$\beta$ run but a small speed-up factor $f_{\rm sp}=6$, as listed in \tabref{tab:list}. Similarly, the z$\beta$\_$t3$ run uses a further smaller speed-up factor $f_{\rm sp}=3$. The z$\beta$\_large run has a mesh of $256\times256\times1080$ with a horizontal/vertical grid spacing of 1080/128 km, such that the domain represents the same horizontal area of the actual active region. We use a lower resolution for this run to save computational resources. This run uses $f_{\rm sp}=12$. Moreover, from the snapshot of the z$\beta$\_large run at 130000 iterations, which corresponds to approximately February 13 6$^{\rm h}$ UT, we start the z$\beta$\_large\_$t3$ run, which uses the same domain size but a much smaller speed-up factor $f_{\rm sp}=3$. The run with a small $f_{\rm sp}$ becomes significantly more expensive. The option of restarting from a snapshot of a large $f_{\rm sp}$ run skips the early evolution and covers the fast emergence stage and the eruption.
 
We compare the total and free magnetic energy of these runs in \figref{fig:emag_scaling}. The speed and volume factors are listed in the legend for each case. All the cases show a quantitatively consistent trend following the emergence of the active region over an equivalent time of 4.5 solar days. A factor that causes the differences between the curves of difference cases is some weak activities that occur before February 15. However, the slopes of the curves, which describe the increasing rate of the magnetic energy, are consistent across all the runs. This is demonstrated in the lower panel of \figref{fig:emag_scaling}, where the time derivatives of the curves in the upper panel are shown following the same line styles.

The consistency of the results from runs with and without a scaling in time and space can also be examined by comparing the zero-$\beta$ run and the MHD evo. run that are already shown in \figref{fig:mag}. In \figref{fig:emag_zbevo}, we compare the magnetic energy of the zero-$\beta$ run, which is scaled by the corresponding factors $f_{\rm sp}=12$ and $f_{\rm V}=1.875^{3}$, and that of the MHD evo. run, which has the same domain size as the actual active region and is driven by the magnetic field at the original cadence. 

The small difference in the total magnetic energy curves is caused by that the potential part of the initial condition of the MHD evo. run is directly calculated from the observed magnetic field (see \sectref{sec:method}), which gives the MHD evo. run a slightly higher starting point. In comparison, the difference between the $E_{\rm free}$ curves is approximately 1\% or smaller. 

The curves of magnetic energy in both run exhibit a consistent trend for more than 2 hours until $t\approx-3000$\,s. The gentle decreasing trend in the zero-$\beta$ run curves (dotted lines) indicates that the eruption is about to occur, whereas the eruption in the MHD evo. run some minutes later ($t\approx200$\,s).

\bibliography{reference}{}

\begin{thebibliography}{}
\expandafter\ifx\csname natexlab\endcsname\relax\def\natexlab#1{#1}\fi
\providecommand{\url}[1]{\href{#1}{#1}}
\providecommand{\dodoi}[1]{doi:~\href{http://doi.org/#1}{\nolinkurl{#1}}}
\providecommand{\doeprint}[1]{\href{http://ascl.net/#1}{\nolinkurl{http://ascl.net/#1}}}
\providecommand{\doarXiv}[1]{\href{https://arxiv.org/abs/#1}{\nolinkurl{https://arxiv.org/abs/#1}}}

\bibitem[{A.~N. {Afanasyev} {et~al.}(2023){Afanasyev}, {Fan}, {Kazachenko}, \&
  {Cheung}}]{Afanasyev+al:2023}
{Afanasyev}, A.~N., {Fan}, Y., {Kazachenko}, M.~D., \& {Cheung}, M. C.~M. 2023,
  \bibinfo{title}{{Hybrid Data-driven Magnetofrictional and Magnetohydrodynamic
  Simulations of an Eruptive Solar Active Region},} \apj, 952, 136,
  \dodoi{10.3847/1538-4357/acd7e9}

\bibitem[{J.~C. {Allred} {et~al.}(2015){Allred}, {Kowalski}, \&
  {Carlsson}}]{Allred+al:2015}
{Allred}, J.~C., {Kowalski}, A.~F., \& {Carlsson}, M. 2015, \bibinfo{title}{{A
  Unified Computational Model for Solar and Stellar Flares},} \apj, 809, 104,
  \dodoi{10.1088/0004-637X/809/1/104}

\bibitem[{M.~J. {Aschwanden} {et~al.}(2016){Aschwanden}, {Holman},
  {O'Flannagain}, {Caspi}, {McTiernan}, \& {Kontar}}]{Aschwanden+al:2016_III}
{Aschwanden}, M.~J., {Holman}, G., {O'Flannagain}, A., {et~al.} 2016,
  \bibinfo{title}{{Global Energetics of Solar Flares. III. Nonthermal
  Energies},} \apj, 832, 27, \dodoi{10.3847/0004-637X/832/1/27}

\bibitem[{F. {Chen}(2025){Chen}}]{Chen:2025}
{Chen}, F. 2025, \bibinfo{title}{{Data-driven Radiative Magnetohydrodynamics
  Simulations with the MURaM code: the Emerging Active Region Corona},} arXiv
  e-prints, arXiv:2511.02362, \dodoi{10.48550/arXiv.2511.02362}

\bibitem[{F. {Chen} {et~al.}(2023{\natexlab{a}}){Chen}, {Cheung}, {Rempel}, \&
  {Chintzoglou}}]{Chen+al:2023}
{Chen}, F., {Cheung}, M. C.~M., {Rempel}, M., \& {Chintzoglou}, G.
  2023{\natexlab{a}}, \bibinfo{title}{{Data-driven Radiative
  Magnetohydrodynamics Simulations with the MURaM Code},} \apj, 949, 118,
  \dodoi{10.3847/1538-4357/acc8c5}

\bibitem[{F. {Chen} {et~al.}(2022){Chen}, {Rempel}, \& {Fan}}]{Chen+al:2022}
{Chen}, F., {Rempel}, M., \& {Fan}, Y. 2022, \bibinfo{title}{{A Comprehensive
  Radiative Magnetohydrodynamics Simulation of Active Region Scale Flux
  Emergence from the Convection Zone to the Corona},} \apj, 937, 91,
  \dodoi{10.3847/1538-4357/ac8f95}

\bibitem[{F. {Chen} {et~al.}(2023{\natexlab{b}}){Chen}, {Rempel}, \&
  {Fan}}]{Chen+al:2023L}
{Chen}, F., {Rempel}, M., \& {Fan}, Y. 2023{\natexlab{b}},
  \bibinfo{title}{{Eruption of a Magnetic Flux Rope in a Comprehensive
  Radiative Magnetohydrodynamic Simulation of Flare-productive Active
  Regions},} \apjl, 950, L3, \dodoi{10.3847/2041-8213/acda2e}

\bibitem[{X. {Cheng} {et~al.}(2020){Cheng}, {Zhang}, {Kliem}, {T{\"o}r{\"o}k},
  {Xing}, {Zhou}, {Inhester}, \& {Ding}}]{ChengXin+al:2020}
{Cheng}, X., {Zhang}, J., {Kliem}, B., {et~al.} 2020,
  \bibinfo{title}{{Initiation and Early Kinematic Evolution of Solar
  Eruptions},} \apj, 894, 85, \dodoi{10.3847/1538-4357/ab886a}

\bibitem[{M.~C.~M. {Cheung} \& M.~L. {DeRosa}(2012){Cheung} \&
  {DeRosa}}]{Cheung+DeRosa:2012}
{Cheung}, M. C.~M., \& {DeRosa}, M.~L. 2012, \bibinfo{title}{{A Method for
  Data-driven Simulations of Evolving Solar Active Regions},} \apj, 757, 147,
  \dodoi{10.1088/0004-637X/757/2/147}

\bibitem[{M.~C.~M. {Cheung} {et~al.}(2019){Cheung}, {Rempel}, {Chintzoglou},
  {Chen}, {Testa}, {Mart{\'\i}nez-Sykora}, {Sainz Dalda}, {DeRosa},
  {Malanushenko}, {Hansteen}, {De Pontieu}, {Carlsson}, {Gudiksen}, \&
  {McIntosh}}]{Cheung+al:2019}
{Cheung}, M.~C.~M., {Rempel}, M., {Chintzoglou}, G., {et~al.} 2019,
  \bibinfo{title}{{A comprehensive three-dimensional radiative
  magnetohydrodynamic simulation of a solar flare},} Nature Astronomy, 3, 160,
  \dodoi{10.1038/s41550-018-0629-3}

\bibitem[{G. {Chintzoglou} {et~al.}(2019){Chintzoglou}, {Zhang}, {Cheung}, \&
  {Kazachenko}}]{Chintzoglou+al:2019}
{Chintzoglou}, G., {Zhang}, J., {Cheung}, M. C.~M., \& {Kazachenko}, M. 2019,
  \bibinfo{title}{{The Origin of Major Solar Activity: Collisional Shearing
  between Nonconjugated Polarities of Multiple Bipoles Emerging within Active
  Regions},} \apj, 871, 67, \dodoi{10.3847/1538-4357/aaef30}

\bibitem[{K. {Dissauer} {et~al.}(2018){Dissauer}, {Veronig}, {Temmer},
  {Podladchikova}, \& {Vanninathan}}]{Dissauer+al:2018}
{Dissauer}, K., {Veronig}, A.~M., {Temmer}, M., {Podladchikova}, T., \&
  {Vanninathan}, K. 2018, \bibinfo{title}{{On the Detection of Coronal Dimmings
  and the Extraction of Their Characteristic Properties},} \apj, 855, 137,
  \dodoi{10.3847/1538-4357/aaadb5}

\bibitem[{Y. {Fan} {et~al.}(2024){Fan}, {Kazachenko}, {Afanasyev}, \&
  {Fisher}}]{Fan+al:2024}
{Fan}, Y., {Kazachenko}, M.~D., {Afanasyev}, A.~N., \& {Fisher}, G.~H. 2024,
  \bibinfo{title}{{A Data-driven Magnetohydrodynamic Simulation of the 2011
  February 15 Coronal Mass Ejection from Active Region NOAA 11158},} \apj, 975,
  206, \dodoi{10.3847/1538-4357/ad7f53}

\bibitem[{G.~H. {Fisher} {et~al.}(1985){Fisher}, {Canfield}, \&
  {McClymont}}]{Fisher+al:1985}
{Fisher}, G.~H., {Canfield}, R.~C., \& {McClymont}, A.~N. 1985,
  \bibinfo{title}{{Flare loop radiative hydrodynamics. V - Response to
  thick-target heating. VI - Chromospheric evaporation due to heating by
  nonthermal electrons. VII - Dynamics of the thick-target heated
  chromosphere},} \apj, 289, 414, \dodoi{10.1086/162901}

\bibitem[{G.~H. {Fisher} {et~al.}(2020){Fisher}, {Kazachenko}, {Welsch}, {Sun},
  {Lumme}, {Bercik}, {DeRosa}, \& {Cheung}}]{Fisher+al:2020}
{Fisher}, G.~H., {Kazachenko}, M.~D., {Welsch}, B.~T., {et~al.} 2020,
  \bibinfo{title}{{The PDFI\_SS Electric Field Inversion Software},} \apjs,
  248, 2, \dodoi{10.3847/1538-4365/ab8303}

\bibitem[{Y. {Guo} {et~al.}(2024){Guo}, {Guo}, {Ni}, {Xia}, {Zhong}, {Ding},
  {Chen}, \& {Keppens}}]{GuoYang+al:2024}
{Guo}, Y., {Guo}, J., {Ni}, Y., {et~al.} 2024, \bibinfo{title}{{Magnetic flux
  rope models and data-driven magnetohydrodynamic simulations of solar
  eruptions},} Reviews of Modern Plasma Physics, 8, 29,
  \dodoi{10.1007/s41614-024-00167-2}

\bibitem[{K. {Hayashi} {et~al.}(2018){Hayashi}, {Feng}, {Xiong}, \&
  {Jiang}}]{Hayashi+al:2018}
{Hayashi}, K., {Feng}, X., {Xiong}, M., \& {Jiang}, C. 2018,
  \bibinfo{title}{{An MHD Simulation of Solar Active Region 11158 Driven with a
  Time-dependent Electric Field Determined from HMI Vector Magnetic Field
  Measurement Data},} \apj, 855, 11, \dodoi{10.3847/1538-4357/aaacd8}

\bibitem[{K. {Hayashi} {et~al.}(2019){Hayashi}, {Feng}, {Xiong}, \&
  {Jiang}}]{Hayashi+al:2019}
{Hayashi}, K., {Feng}, X., {Xiong}, M., \& {Jiang}, C. 2019,
  \bibinfo{title}{{Magnetohydrodynamic Simulations for Solar Active Regions
  using Time-series Data of Surface Plasma Flow and Electric Field Inferred
  from Helioseismic Magnetic Imager Vector Magnetic Field Measurements},}
  \apjl, 871, L28, \dodoi{10.3847/2041-8213/aaffcf}

\bibitem[{J.~T. {Hoeksema} {et~al.}(2014){Hoeksema}, {Liu}, {Hayashi}, {Sun},
  {Schou}, {Couvidat}, {Norton}, {Bobra}, {Centeno}, {Leka}, {Barnes}, \&
  {Turmon}}]{Hoeksema+al:2014}
{Hoeksema}, J.~T., {Liu}, Y., {Hayashi}, K., {et~al.} 2014,
  \bibinfo{title}{{The Helioseismic and Magnetic Imager (HMI) Vector Magnetic
  Field Pipeline: Overview and Performance},} \solphys, 289, 3483,
  \dodoi{10.1007/s11207-014-0516-8}

\bibitem[{J.~T. {Hoeksema} {et~al.}(2020){Hoeksema}, {Abbett}, {Bercik},
  {Cheung}, {DeRosa}, {Fisher}, {Hayashi}, {Kazachenko}, {Liu}, {Lumme},
  {Lynch}, {Sun}, \& {Welsch}}]{Hoeksema+al:2020}
{Hoeksema}, J.~T., {Abbett}, W.~P., {Bercik}, D.~J., {et~al.} 2020,
  \bibinfo{title}{{The Coronal Global Evolutionary Model: Using HMI Vector
  Magnetogram and Doppler Data to Determine Coronal Magnetic Field Evolution},}
  \apjs, 250, 28, \dodoi{10.3847/1538-4365/abb3fb}

\bibitem[{S. {Inoue} {et~al.}(2014){Inoue}, {Hayashi}, {Magara}, {Choe}, \&
  {Park}}]{Inoue+al:2014}
{Inoue}, S., {Hayashi}, K., {Magara}, T., {Choe}, G.~S., \& {Park}, Y.~D. 2014,
  \bibinfo{title}{{Magnetohydrodynamic Simulation of the X2.2 Solar Flare on
  2011 February 15. I. Comparison with the Observations},} \apj, 788, 182,
  \dodoi{10.1088/0004-637X/788/2/182}

\bibitem[{S. {Inoue} {et~al.}(2015){Inoue}, {Hayashi}, {Magara}, {Choe}, \&
  {Park}}]{Inoue+al:2015}
{Inoue}, S., {Hayashi}, K., {Magara}, T., {Choe}, G.~S., \& {Park}, Y.~D. 2015,
  \bibinfo{title}{{Magnetohydrodynamic Simulation of the X2.2 Solar Flare on
  2011 February 15. II. Dynamics Connecting the Solar Flare and the Coronal
  Mass Ejection},} \apj, 803, 73, \dodoi{10.1088/0004-637X/803/2/73}

\bibitem[{S. {Inoue} {et~al.}(2018){Inoue}, {Kusano}, {B{\"u}chner}, \&
  {Sk{\'a}la}}]{Inoue+al:2018}
{Inoue}, S., {Kusano}, K., {B{\"u}chner}, J., \& {Sk{\'a}la}, J. 2018,
  \bibinfo{title}{{Formation and dynamics of a solar eruptive flux tube},}
  Nature Communications, 9, 174, \dodoi{10.1038/s41467-017-02616-8}

\bibitem[{M. {Janvier} {et~al.}(2014){Janvier}, {Aulanier}, {Bommier},
  {Schmieder}, {D{\'e}moulin}, \& {Pariat}}]{Janvier+al:2014}
{Janvier}, M., {Aulanier}, G., {Bommier}, V., {et~al.} 2014,
  \bibinfo{title}{{Electric Currents in Flare Ribbons: Observations and
  Three-dimensional Standard Model},} \apj, 788, 60,
  \dodoi{10.1088/0004-637X/788/1/6010.48550/arXiv.1402.2010}

\bibitem[{R. {Jarolim} {et~al.}(2023){Jarolim}, {Thalmann}, {Veronig}, \&
  {Podladchikova}}]{Jarolim+al:2023}
{Jarolim}, R., {Thalmann}, J.~K., {Veronig}, A.~M., \& {Podladchikova}, T.
  2023, \bibinfo{title}{{Probing the solar coronal magnetic field with
  physics-informed neural networks.},} Nature Astronomy, 7, 1171,
  \dodoi{10.1038/s41550-023-02030-9}

\bibitem[{C. {Jiang}(2024){Jiang}}]{JiangChaowei:2024}
{Jiang}, C. 2024, \bibinfo{title}{{From fundamental theory to realistic
  modeling of the birth of solar eruptions},} Science China Earth Sciences, 67,
  3765, \dodoi{10.1007/s11430-023-1402-3}

\bibitem[{C. {Jiang} {et~al.}(2022){Jiang}, {Feng}, {Guo}, \&
  {Hu}}]{JiangChaowei+al:2022}
{Jiang}, C., {Feng}, X., {Guo}, Y., \& {Hu}, Q. 2022,
  \bibinfo{title}{{Data-driven modeling of solar coronal magnetic field
  evolution and eruptions},} The Innovation, 3, 100236,
  \dodoi{10.1016/j.xinn.2022.100236}

\bibitem[{C. {Jiang} {et~al.}(2021){Jiang}, {Feng}, {Liu}, {Yan}, {Hu},
  {Moore}, {Duan}, {Cui}, {Zuo}, {Wang}, \& {Wei}}]{JiangChaowei+al:2021}
{Jiang}, C., {Feng}, X., {Liu}, R., {et~al.} 2021, \bibinfo{title}{{A
  fundamental mechanism of solar eruption initiation},} Nature Astronomy, 5,
  1126, \dodoi{10.1038/s41550-021-01414-z}

\bibitem[{M. {Jin} {et~al.}(2022){Jin}, {Cheung}, {DeRosa}, {Nitta}, \&
  {Schrijver}}]{Jin+al:2022}
{Jin}, M., {Cheung}, M. C.~M., {DeRosa}, M.~L., {Nitta}, N.~V., \& {Schrijver},
  C.~J. 2022, \bibinfo{title}{{Coronal Mass Ejections and Dimmings: A
  Comparative Study Using MHD Simulations and SDO Observations},} \apj, 928,
  154, \dodoi{10.3847/1538-4357/ac589b}

\bibitem[{M.~D. {Kazachenko} {et~al.}(2014){Kazachenko}, {Fisher}, \&
  {Welsch}}]{Kazachenko+al:2014}
{Kazachenko}, M.~D., {Fisher}, G.~H., \& {Welsch}, B.~T. 2014,
  \bibinfo{title}{{A Comprehensive Method of Estimating Electric Fields from
  Vector Magnetic Field and Doppler Measurements},} \apj, 795, 17,
  \dodoi{10.1088/0004-637X/795/1/17}

\bibitem[{M.~D. {Kazachenko} {et~al.}(2015){Kazachenko}, {Fisher}, {Welsch},
  {Liu}, \& {Sun}}]{Kazachenko+al:2015}
{Kazachenko}, M.~D., {Fisher}, G.~H., {Welsch}, B.~T., {Liu}, Y., \& {Sun}, X.
  2015, \bibinfo{title}{{Photospheric Electric Fields and Energy Fluxes in the
  Eruptive Active Region NOAA 11158},} \apj, 811, 16,
  \dodoi{10.1088/0004-637X/811/1/16}

\bibitem[{A.~G. {Kosovichev}(2011){Kosovichev}}]{Kosovichev:2011}
{Kosovichev}, A.~G. 2011, \bibinfo{title}{{Helioseismic Response to the X2.2
  Solar Flare of 2011 February 15},} \apjl, 734, L15,
  \dodoi{10.1088/2041-8205/734/1/L15}

\bibitem[{A.~F. {Kowalski} {et~al.}(2017){Kowalski}, {Allred}, {Daw}, {Cauzzi},
  \& {Carlsson}}]{Kowalski+al:2017}
{Kowalski}, A.~F., {Allred}, J.~C., {Daw}, A., {Cauzzi}, G., \& {Carlsson}, M.
  2017, \bibinfo{title}{{The Atmospheric Response to High Nonthermal Electron
  Beam Fluxes in Solar Flares. I. Modeling the Brightest NUV Footpoints in the
  X1 Solar Flare of 2014 March 29},} \apj, 836, 12,
  \dodoi{10.3847/1538-4357/836/1/12}

\bibitem[{S. Li {et~al.}(2019)Li, Jaroszynski, Pearse, Orf, \& Clyne}]{vapor}
Li, S., Jaroszynski, S., Pearse, S., Orf, L., \& Clyne, J. 2019,
  \bibinfo{title}{VAPOR: A Visualization Package Tailored to Analyze Simulation
  Data in Earth System Science,} Atmosphere, 10, \dodoi{10.3390/atmos10090488}

\bibitem[{Y. {Liu} \& P.~W. {Schuck}(2012){Liu} \& {Schuck}}]{Liu+Schuck:2012}
{Liu}, Y., \& {Schuck}, P.~W. 2012, \bibinfo{title}{{Magnetic Energy and
  Helicity in Two Emerging Active Regions in the Sun},} \apj, 761, 105,
  \dodoi{10.1088/0004-637X/761/2/105}

\bibitem[{E. {Lumme} {et~al.}(2017){Lumme}, {Pomoell}, \&
  {Kilpua}}]{Lumme+al:2017}
{Lumme}, E., {Pomoell}, J., \& {Kilpua}, E.~K.~J. 2017,
  \bibinfo{title}{{Optimization of Photospheric Electric Field Estimates for
  Accurate Retrieval of Total Magnetic Energy Injection},} \solphys, 292, 191,
  \dodoi{10.1007/s11207-017-1214-0}

\bibitem[{R.~O. {Milligan} {et~al.}(2014){Milligan}, {Kerr}, {Dennis},
  {Hudson}, {Fletcher}, {Allred}, {Chamberlin}, {Ireland}, {Mathioudakis}, \&
  {Keenan}}]{Milligan+al:2014}
{Milligan}, R.~O., {Kerr}, G.~S., {Dennis}, B.~R., {et~al.} 2014,
  \bibinfo{title}{{The Radiated Energy Budget of Chromospheric Plasma in a
  Major Solar Flare Deduced from Multi-wavelength Observations},} \apj, 793,
  70, \dodoi{10.1088/0004-637X/793/2/70}

\bibitem[{G.~E. {Moreton} \& H.~E. {Ramsey}(1960){Moreton} \&
  {Ramsey}}]{Moreton+Ramsey:1960}
{Moreton}, G.~E., \& {Ramsey}, H.~E. 1960, \bibinfo{title}{{Recent Observations
  of Dynamical Phenomena Associated with Solar Flares},} \pasp, 72, 357,
  \dodoi{10.1086/127549}

\bibitem[{S. {Patsourakos} {et~al.}(2020){Patsourakos}, {Vourlidas},
  {T{\"o}r{\"o}k}, {Kliem}, {Antiochos}, {Archontis}, {Aulanier}, {Cheng},
  {Chintzoglou}, {Georgoulis}, {Green}, {Leake}, {Moore}, {Nindos}, {Syntelis},
  {Yardley}, {Yurchyshyn}, \& {Zhang}}]{Patsourakos+al:2020}
{Patsourakos}, S., {Vourlidas}, A., {T{\"o}r{\"o}k}, T., {et~al.} 2020,
  \bibinfo{title}{{Decoding the Pre-Eruptive Magnetic Field Configurations of
  Coronal Mass Ejections},} \ssr, 216, 131, \dodoi{10.1007/s11214-020-00757-9}

\bibitem[{J.~W. {Reep} {et~al.}(2015){Reep}, {Bradshaw}, \&
  {Alexander}}]{Reep+al:2015}
{Reep}, J.~W., {Bradshaw}, S.~J., \& {Alexander}, D. 2015,
  \bibinfo{title}{{Optimal Electron Energies for Driving Chromospheric
  Evaporation in Solar Flares},} \apj, 808, 177,
  \dodoi{10.1088/0004-637X/808/2/177}

\bibitem[{M. {Rempel} {et~al.}(2023){Rempel}, {Chintzoglou}, {Cheung}, {Fan},
  \& {Kleint}}]{Rempel+al:2023}
{Rempel}, M., {Chintzoglou}, G., {Cheung}, M. C.~M., {Fan}, Y., \& {Kleint}, L.
  2023, \bibinfo{title}{{Comprehensive Radiative MHD Simulations of Eruptive
  Flares above Collisional Polarity Inversion Lines},} \apj, 955, 105,
  \dodoi{10.3847/1538-4357/aced4d}

\bibitem[{A. {Savcheva} {et~al.}(2015){Savcheva}, {Pariat}, {McKillop},
  {McCauley}, {Hanson}, {Su}, {Werner}, \& {DeLuca}}]{Savcheva+al:2015}
{Savcheva}, A., {Pariat}, E., {McKillop}, S., {et~al.} 2015,
  \bibinfo{title}{{The Relation between Solar Eruption Topologies and Observed
  Flare Features. I. Flare Ribbons},} \apj, 810, 96,
  \dodoi{10.1088/0004-637X/810/2/96}

\bibitem[{P.~H. {Scherrer} {et~al.}(2012){Scherrer}, {Schou}, {Bush},
  {Kosovichev}, {Bogart}, {Hoeksema}, {Liu}, {Duvall}, {Zhao}, {Title},
  {Schrijver}, {Tarbell}, \& {Tomczyk}}]{HMI}
{Scherrer}, P.~H., {Schou}, J., {Bush}, R.~I., {et~al.} 2012,
  \bibinfo{title}{{The Helioseismic and Magnetic Imager (HMI) Investigation for
  the Solar Dynamics Observatory (SDO)},} \solphys, 275, 207,
  \dodoi{10.1007/s11207-011-9834-2}

\bibitem[{B. {Schmieder} {et~al.}(2024){Schmieder}, {Guo}, \&
  {Poedts}}]{Schmieder+al:2024}
{Schmieder}, B., {Guo}, J., \& {Poedts}, S. 2024, \bibinfo{title}{{Recent
  advances in solar data-driven MHD simulations of the formation and evolution
  of CME flux ropes},} Reviews of Modern Plasma Physics, 8, 27,
  \dodoi{10.1007/s41614-024-00166-3}

\bibitem[{C.~J. {Schrijver} {et~al.}(2011){Schrijver}, {Aulanier}, {Title},
  {Pariat}, \& {Delann{\'e}e}}]{Schrijver+al:2011}
{Schrijver}, C.~J., {Aulanier}, G., {Title}, A.~M., {Pariat}, E., \&
  {Delann{\'e}e}, C. 2011, \bibinfo{title}{{The 2011 February 15 X2 Flare,
  Ribbons, Coronal Front, and Mass Ejection: Interpreting the Three-dimensional
  Views from the Solar Dynamics Observatory and STEREO Guided by
  Magnetohydrodynamic Flux-rope Modeling},} \apj, 738, 167,
  \dodoi{10.1088/0004-637X/738/2/167}

\bibitem[{X. {Sun} {et~al.}(2017){Sun}, {Hoeksema}, {Liu}, {Kazachenko}, \&
  {Chen}}]{SunXudong+al:2017}
{Sun}, X., {Hoeksema}, J.~T., {Liu}, Y., {Kazachenko}, M., \& {Chen}, R. 2017,
  \bibinfo{title}{{Investigating the Magnetic Imprints of Major Solar Eruptions
  with SDO/HMI High-cadence Vector Magnetograms},} \apj, 839, 67,
  \dodoi{10.3847/1538-4357/aa69c1}

\bibitem[{X. {Sun} {et~al.}(2012){Sun}, {Hoeksema}, {Liu}, {Wiegelmann},
  {Hayashi}, {Chen}, \& {Thalmann}}]{SunXudong+al:2012}
{Sun}, X., {Hoeksema}, J.~T., {Liu}, Y., {et~al.} 2012,
  \bibinfo{title}{{Evolution of Magnetic Field and Energy in a Major Eruptive
  Active Region Based on SDO/HMI Observation},} \apj, 748, 77,
  \dodoi{10.1088/0004-637X/748/2/77}

\bibitem[{L.~A. {Tarr} {et~al.}(2024){Tarr}, {Kee}, {Linton}, {Schuck}, \&
  {Leake}}]{Tarr+al:2024}
{Tarr}, L.~A., {Kee}, N.~D., {Linton}, M.~G., {Schuck}, P.~W., \& {Leake},
  J.~E. 2024, \bibinfo{title}{{Simulating the Photospheric to Coronal Plasma
  Using Magnetohydrodynamic Characteristics. I. Data-driven Boundary
  Conditions},} \apjs, 270, 30, \dodoi{10.3847/1538-4365/ad0e0c}

\bibitem[{S. {Toriumi} {et~al.}(2017){Toriumi}, {Schrijver}, {Harra}, {Hudson},
  \& {Nagashima}}]{Toriumi+al:2017b}
{Toriumi}, S., {Schrijver}, C.~J., {Harra}, L.~K., {Hudson}, H., \&
  {Nagashima}, K. 2017, \bibinfo{title}{{Magnetic Properties of Solar Active
  Regions That Govern Large Solar Flares and Eruptions},} \apj, 834, 56,
  \dodoi{10.3847/1538-4357/834/1/56}

\bibitem[{K. {Tziotziou} {et~al.}(2013){Tziotziou}, {Georgoulis}, \&
  {Liu}}]{Tziotziou+al:2013}
{Tziotziou}, K., {Georgoulis}, M.~K., \& {Liu}, Y. 2013,
  \bibinfo{title}{{Interpreting Eruptive Behavior in NOAA AR 11158 via the
  Region's Magnetic Energy and Relative-helicity Budgets},} \apj, 772, 115,
  \dodoi{10.1088/0004-637X/772/2/115}

\bibitem[{Y. {Uchida}(1968){Uchida}}]{Uchida:1968}
{Uchida}, Y. 1968, \bibinfo{title}{{Propagation of Hydromagnetic Disturbances
  in the Solar Corona and Moreton's Wave Phenomenon},} \solphys, 4, 30,
  \dodoi{10.1007/BF00146996}

\bibitem[{P. {Vemareddy} {et~al.}(2012){Vemareddy}, {Ambastha}, \&
  {Maurya}}]{Vemareddy+al:2012}
{Vemareddy}, P., {Ambastha}, A., \& {Maurya}, R.~A. 2012, \bibinfo{title}{{On
  the Role of Rotating Sunspots in the Activity of Solar Active Region NOAA
  11158},} \apj, 761, 60, \dodoi{10.1088/0004-637X/761/1/60}

\bibitem[{C. {Wang} {et~al.}(2023){Wang}, {Chen}, {Ding}, \&
  {Lu}}]{WangCan+al:2023}
{Wang}, C., {Chen}, F., {Ding}, M., \& {Lu}, Z. 2023,
  \bibinfo{title}{{Radiative Magnetohydrodynamic Simulation of the Confined
  Eruption of a Magnetic Flux Rope: Unveiling the Driving and Constraining
  Forces},} \apj, 956, 106, \dodoi{10.3847/1538-4357/acedfe}

\bibitem[{S. {Wang} {et~al.}(2012){Wang}, {Liu}, {Liu}, {Deng}, {Liu}, \&
  {Wang}}]{WangShuo+al:2012}
{Wang}, S., {Liu}, C., {Liu}, R., {et~al.} 2012, \bibinfo{title}{{Response of
  the Photospheric Magnetic Field to the X2.2 Flare on 2011 February 15},}
  \apjl, 745, L17, \dodoi{10.1088/2041-8205/745/2/L17}

\bibitem[{A. {Warmuth} \& G. {Mann}(2016){Warmuth} \&
  {Mann}}]{Warmuth+Mann:2016b}
{Warmuth}, A., \& {Mann}, G. 2016, \bibinfo{title}{{Constraints on energy
  release in solar flares from RHESSI and GOES X-ray observations. II.
  Energetics and energy partition},} \aap, 588, A116,
  \dodoi{10.1051/0004-6361/201527475}

\bibitem[{P.~R. {Young} {et~al.}(2013){Young}, {Doschek}, {Warren}, \&
  {Hara}}]{Young+al:2013}
{Young}, P.~R., {Doschek}, G.~A., {Warren}, H.~P., \& {Hara}, H. 2013,
  \bibinfo{title}{{Properties of a Solar Flare Kernel Observed by Hinode and
  SDO},} \apj, 766, 127, \dodoi{10.1088/0004-637X/766/2/127}

\bibitem[{S. {Zharkov} {et~al.}(2011){Zharkov}, {Green}, {Matthews}, \&
  {Zharkova}}]{Zharkov+al:2011}
{Zharkov}, S., {Green}, L.~M., {Matthews}, S.~A., \& {Zharkova}, V.~V. 2011,
  \bibinfo{title}{{2011 February 15: Sunquakes Produced by Flux Rope
  Eruption},} \apjl, 741, L35, \dodoi{10.1088/2041-8205/741/2/L35}

\end{thebibliography}
\bibliographystyle{aasjournalv7}

\end{document}